# Microarrays, Empirical Bayes and the Two-Groups Model

**Bradley Efron**

*Abstract.* The classic frequentist theory of hypothesis testing developed by Neyman, Pearson and Fisher has a claim to being the twentieth century's most influential piece of applied mathematics. Something new is happening in the twenty-first century: high-throughput devices, such as microarrays, routinely require simultaneous hypothesis tests for thousands of individual cases, not at all what the classical theory had in mind. In these situations empirical Bayes information begins to force itself upon frequentists and Bayesians alike. The two-groups model is a simple Bayesian construction that facilitates empirical Bayes analysis. This article concerns the interplay of Bayesian and frequentist ideas in the two-groups setting, with particular attention focused on Benjamini and Hochberg's False Discovery Rate method. Topics include the choice and meaning of the null hypothesis in large-scale testing situations, power considerations, the limitations of permutation methods, significance testing for groups of cases (such as pathways in microarray studies), correlation effects, multiple confidence intervals and Bayesian competitors to the two-groups model.

## 1. INTRODUCTION

Simultaneous hypothesis testing was a lively research topic during my student days, exemplified by Rupert Miller's classic text "Simultaneous Statistical Inference" (1966, 1981). Attention focused on testing $N$ null hypotheses at the same time, where $N$ was typically less than half a dozen, though the requisite tables might go up to $N = 20$. Modern scientific technology, led by the microarray, has upped the ante in dramatic fashion: my examples here will have $N$'s ranging from 200 to 10,000, while $N = 500{,}000$, from SNP analyses, is waiting in the wings. [The astrostatistical applications in Liang et al. (2004) envision $N = 10^{10}$ and more!]

Miller's text is relentlessly frequentist, reflecting a classic Neyman–Pearson testing framework, with the main goal being preservation of "$\alpha$," overall test size, in the face of multiple inference. Most of the current microarray statistics literature shares this goal, and also its frequentist viewpoint, as described in the nice review article by Dudoit and Boldrick (2003).

Something changes, though, when $N$ gets big: with thousands of parallel inference problems to consider simultaneously, Bayesian considerations begin to force themselves even upon dedicated frequentists. The "two-groups model" of the title is a particularly simple Bayesian framework for large-scale testing situations. This article explores the interplay of frequentist and Bayesian ideas in the two-groups setting,

*Bradley Efron is Professor, Department of Statistics, Stanford University, Stanford, California 94305, USA e-mail: brad@stat.stanford.edu.*

[1]Discussed in 10.1214/07-STS236B, 10.1214/07-STS236C, 10.1214/07-STS236D and 10.1214/07-STS236A; rejoinder at 10.1214/08-STS236REJ.

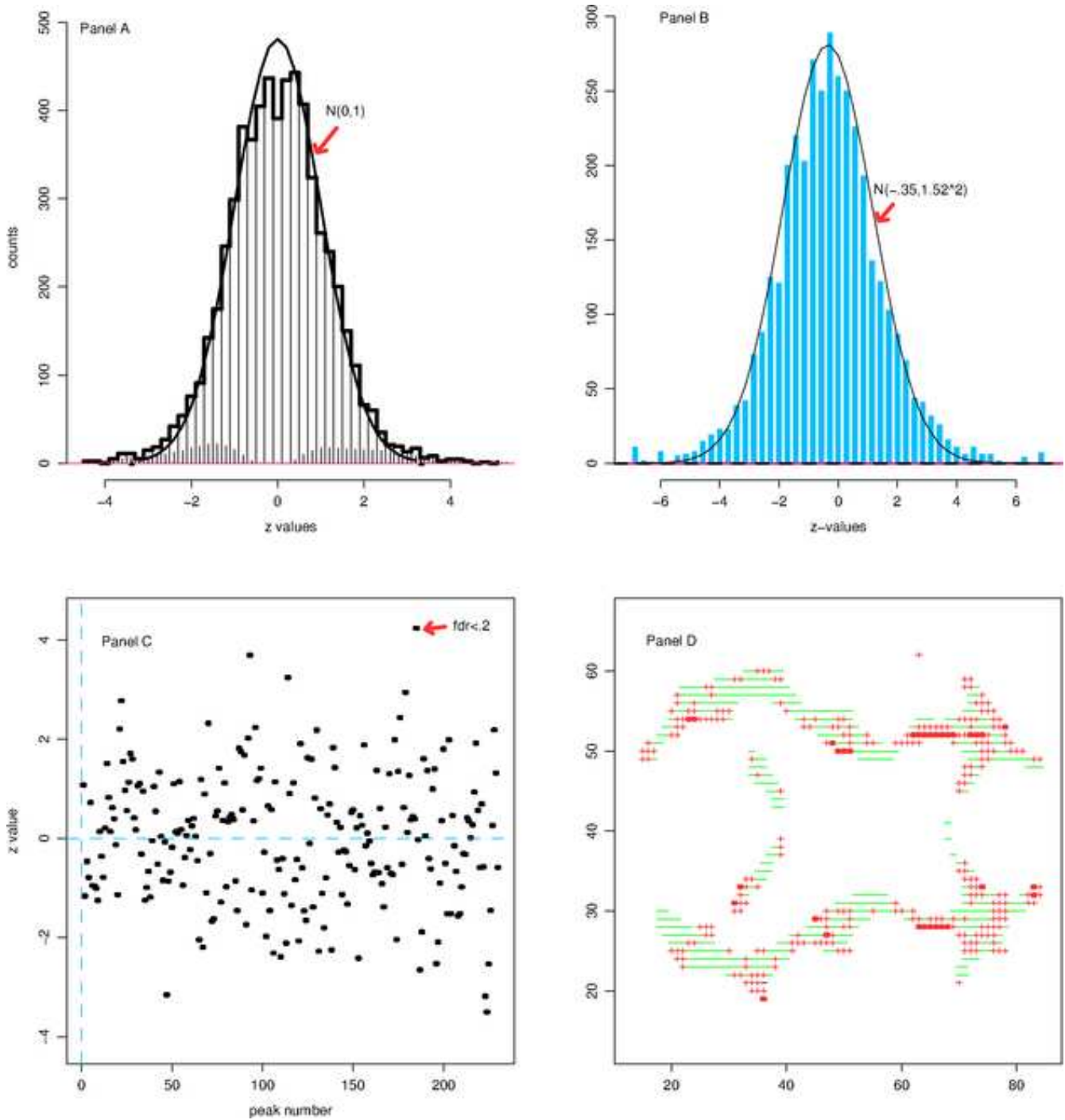

FIG. 1. *Four examples of large-scale simultaneous inference, each panel indicating $N$ z-values as explained in the text. Panel A, prostate cancer microarray study, $N = 6033$ genes; panel B, comparison of advantaged versus disadvantaged students passing mathematics competency tests, $N = 3748$ high schools; panel C, proteomics study, $N = 230$ ordered peaks in time-of-flight spectroscopy experiment; panel D, imaging study comparing dyslexic versus normal children, showing horizontal slice of 655 voxels out of $N = 15{,}455$, coded "$-$" for $z_i < 0$, "$+$" for $z_i \geq 0$ and solid circle for $z_i > 2$.*

with particular attention paid to False Discovery Rates (Benjamini and Hochberg, 1995).

Figure 1 concerns four examples of large-scale simultaneous hypothesis testing. Each example consists of $N$ individual cases, with each case represented by its own $z$-value "$z_i$," for $i = 1, 2, \ldots, N$. The $z_i$'s are based on familiar constructions that, theoretically, should yield standard $N(0, 1)$ normal distributions under a classical null hypothesis,

(1.1)  theoretical null: $z_i \sim N(0, 1)$.

Here is a brief description of the four examples, with further information following as needed in the sequel.

EXAMPLE A [Prostate data, Singh et al. (2002)]. $N = 6033$ genes on 102 microarrays, $n_1 = 50$ healthy males compared with $n_2 = 52$ prostate cancer patients; $z_i$'s based on two-sample $t$ statistics comparing the two categories.

EXAMPLE B [Education data, Rogosa (2003)]. $N = 3748$ California high schools; $z_i$'s based on binomial test of proportion advantaged versus proportion disadvantaged students passing mathematics competency tests.

EXAMPLE C [Proteomics data, Turnbull (2006)]. $N = 230$ ordered peaks in time-of-flight spectroscopy study of 551 heart disease patients. Each peak's $z$-value was obtained from a Cox regression of the patients' survival times, with the predictor variable being the 551 observed intensities at that peak.

EXAMPLE D [Imaging data, Schwartzman et al. (2005)]. $N = 15{,}445$ voxels in a diffusion tensor imaging (DTI) study comparing 6 dyslexic with six normal children; $z_i$'s based on two-sample $t$ statistics comparing the two groups. The figure shows only a single horizontal brain section having 655 voxels, with "−" indicating $z_i < 0$, "+" for $z_i \geq 0$, and solid circles for $z_i > 2$.

Our four examples are enough alike to be usefully analyzed by the two-groups model of Section 2, but there are some striking differences, too: the theoretical $N(0, 1)$ null (1.1) is obviously inappropriate for the education data of panel B; there is a hint of correlation of $z$-value with peak number in panel C, especially near the right limit; and there is substantial spatial correlation appearing in the imaging data of panel D.

My plan here is to discuss a range of inference problems raised by large-scale hypothesis testing, many of which, it seems to me, have been more or less underemphasized in a literature focused on controlling Type-I errors: the choice of a null hypothesis, limitations of permutation methods, the meaning of "null" and "nonnull" in large-scale settings, questions of power, test of significance for groups of cases (e.g., pathways in microarray studies), the effects of correlation, multiple confidence statements and Bayesian competitors to the two-groups model. The presentation is intended to be as nontechnical as possible, many of the topics being discussed more carefully in Efron (2004, 2005, 2006). References will be provided as we go along, but this is not intended as a comprehensive review. Microarrays have stimulated a burst of creativity from the statistics community, and I apologize in advance for this article's concentration on my own point of view, which aims at minimizing the amount of statistical modeling required of the statistician. More model-intensive techniques, including fully Bayesian approaches, as in Parmigiani et al. (2002) or Lewin et al. (2006), have their own virtues, which I hope will emerge in the Discussion.

Section 2 discusses the two-groups model and false discovery rates in an idealized Bayesian setting. Empirical Bayes methods are needed to carry out these ideas in practice, as discussed in Section 3. This discussion assumes a "good" situation, like that of Example A, where the theoretical null (1.1) fits the data. When it does not, as in Example B, the *empirical null* methods of Section 4 come into play. These raise interpretive questions of their own, as mentioned above, discussed in the later sections.

We are living through a scientific revolution powered by the new generation of high-throughput observational devices. This is a wonderful opportunity for statisticians, to redemonstrate our value to the scientific world, but also to rethink basic topics in statistical theory. Hypothesis testing is the topic here, a subject that needs a fresh look in contexts like those of Figure 1.

## 2. THE TWO-GROUPS MODEL AND FALSE DISCOVERY RATES

The two-groups model is too simple to have a single identifiable author, but it plays an important role in the Bayesian microarray literature, as in Lee et al. (2000), Newton et al. (2001) and Efron et al. (2001). We suppose that the $N$ cases ("genes" as they will be called now in deference to microarray studies, though they are not genes in the last three examples of Figure 1) are each either *null* or *nonnull* with prior probability $p_0$ or $p_1 = 1 - p_0$, and with z-values having density either $f_0(z)$ or $f_1(z)$,

$$
\begin{aligned}
p_0 &= \Pr\{\text{null}\} & f_0(z) \text{ density if null}, \\
p_1 &= \Pr\{\text{nonnull}\} & f_1(z) \text{ density if nonnull}.
\end{aligned}
\tag{2.1}
$$

The usual purpose of large-scale simultaneous testing is to reduce a vast set of possibilities to a much smaller set of scientifically interesting prospects. In Example A, for instance, the investigators were probably searching for a few genes, or a few hundred at most, worthy of intensive study for prostate cancer etiology. I will assume

$$p_0 \geq 0.90 \tag{2.2}$$

in what follows, limiting the nonnull genes to no more than 10%.

False discovery rate (Fdr) methods have developed in a strict frequentist framework, beginning with Benjamini and Hochberg's seminal 1995 paper, but they also have a convincing Bayesian rationale in terms of the two-groups model. Let $F_0(z)$ and $F_1(z)$ denote the cumulative distribution functions (cdf) of $f_0(z)$ and $f_1(z)$ in (2.1), and define the mixture cdf $F(z) = p_0 F_0(z) + p_1 F_1(z)$. Then Bayes' rule yields the a posteriori probability of a gene being in the null group of (2.1) given that its z-value $Z$ is less than some threshold $z$, say "Fdr$(z)$," as

$$
\begin{aligned}
\text{Fdr}(z) &\equiv \Pr\{\text{null}|Z \leq z\} \\
&= p_0 F_0(z)/F(z).
\end{aligned}
\tag{2.3}
$$

[Here it is notationally convenient to consider the negative end of the z scale, values like $z = -3$. Definition (2.3) could just as well be changed to $Z > z$ or $Z > |z|$.] Benjamini and Hochberg's (1995) false discovery rate control rule begins by estimating $F(z)$ with the empirical cdf

$$\bar{F}(z) = \#\{z_i \leq z\}/N, \tag{2.4}$$

yielding $\overline{\text{Fdr}}(z) = p_0 F_0(z)/\bar{F}(z)$. The rule selects a control level "$q$," say $q = 0.1$, and then declares as nonnull those genes having z-values $z_i$ satisfying $z_i \leq z_0$, where $z_0$ is the maximum value of $z$ satisfying

$$\overline{\text{Fdr}}(z_0) \leq q \tag{2.5}$$

[usually taking $p_0 = 1$ in (2.3), and $F_0$ the theoretical null, the standard normal cdf $\Phi(z)$ of (1.1)].

The striking theorem proved in the 1995 paper was that the expected proportion of null genes reported by a statistician following rule (2.5) will be no greater than $q$. This assumes independence among the $z_i$'s, extended later to various dependence models in Benjamini and Yekutieli (2001). The theorem is a purely frequentist result, but as pointed out in Storey (2002) and Efron and Tibshirani (2002), it has a simple Bayesian interpretation via (2.3): rule (2.5) is essentially equivalent to declaring nonnull those genes whose estimated tail-area posterior probability of being null is no greater than $q$. It is usually a good sign when Bayesian and frequentist ideas converge on a single methodology, as they do here.

Densities are more natural than tail areas for Bayesian fdr interpretation. Defining the *mixture density* from (2.1),

$$f(z) = p_0 f_0(z) + p_1 f_1(z), \tag{2.6}$$

Bayes' rule gives

$$
\begin{aligned}
\text{fdr}(z) &\equiv \Pr\{\text{null}|Z = z\} \\
&= p_0 f_0(z)/f(z)
\end{aligned}
\tag{2.7}
$$

for the probability of a gene being in the null group given z-score $z$. Here fdr$(z)$ is the *local false discovery rate* (Efron et al., 2001; Efron, 2005).

There is a simple relationship between Fdr$(z)$ and fdr$(z)$,

$$\text{Fdr}(z) = E_f\{\text{fdr}(Z)|Z \leq z\}, \tag{2.8}$$

"$E_f$" indicating expectation with respect to the mixture density $f(z)$. That is, Fdr$(z)$ is the mixture average of fdr$(Z)$ for $Z \leq z$. In the usual situation where fdr$(z)$ decreases as $|z|$ gets large, Fdr$(z)$ will be smaller than fdr$(z)$. Intuitively, if we decide to label all genes with $z_i$ less than some negative value $z_0$ as nonnull, then fdr$(z_0)$, the false discovery rate at the boundary point $z_0$, will be greater than Fdr$(z_0)$, the average false discovery rate beyond the boundary. Figure 2 illustrates the geometrical relationship between Fdr$(z)$ and fdr$(z)$; the Benjamini–Hochberg Fdr control rule amounts to an upper bound on the secant slope.

For Lehmann alternatives

$$F_1(z) = F_0(z)^\gamma, \quad [\gamma < 1], \tag{2.9}$$

it turns out that

$$
\begin{aligned}
&\log\left\{\frac{\text{fdr}(z)}{1-\text{fdr}(z)}\right\} \\
&\quad = \log\left\{\frac{\text{Fdr}(z)}{1-\text{Fdr}(z)}\right\} + \log\left(\frac{1}{\gamma}\right),
\end{aligned}
\tag{2.10}
$$

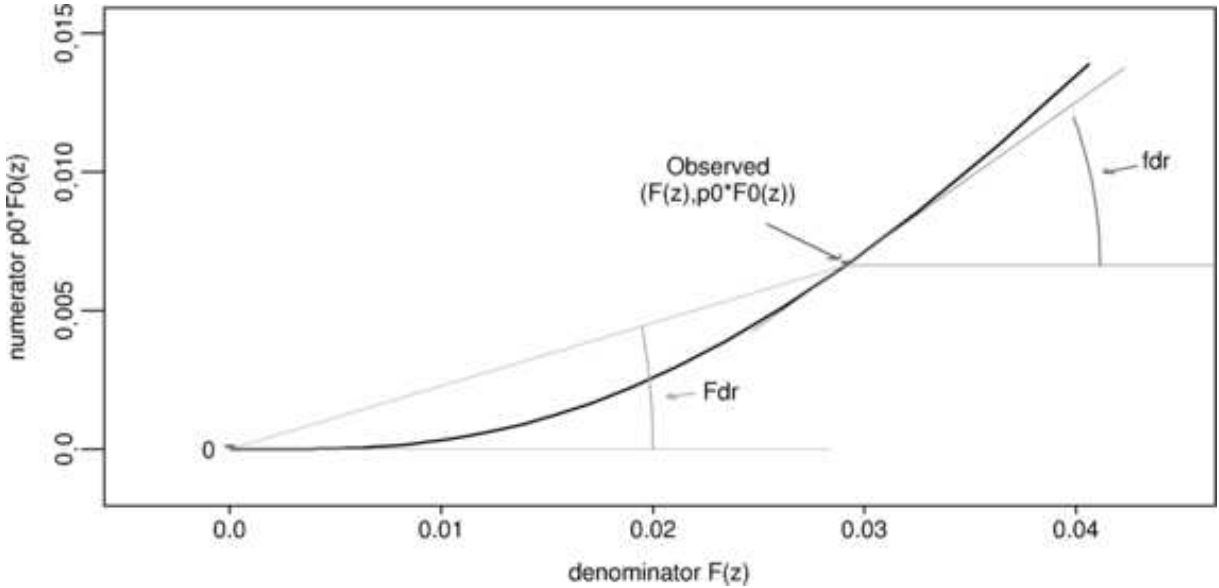

FIG. 2. *Relationship of Fdr(z) to fdr(z). Heavy curve plots numerator of Fdr, $p_0 F_0(z)$, versus denominator $F(z)$; fdr(z) is slope of tangent, Fdr slope of secant.*

so

$$\text{(2.11)} \qquad \text{fdr}(z) \doteq \text{Fdr}(z)/\gamma$$

for small values of Fdr. The prostate data of Figure 1 has $\gamma$ about 1/2 in each tail, making $\text{fdr}(z) \sim 2\,\text{Fdr}(z)$ near the extremes.

The statistics literature has not reached consensus on the choice of $q$ for the Benjamini–Hochberg control rule (2.5)—what would be the equivalent of 0.05 for classical testing—but Bayes factor calculations offer some insight. Efron (2005, 2006) uses the cutoff point

$$\text{(2.12)} \qquad \text{fdr}(z) \leq 0.20$$

for reporting nonnull genes, on the admittedly subjective grounds that fdr values much greater than 0.20 are dangerously prone to wasting investigators' resources. Then (2.6), (2.7) yield posterior odds ratio

$$\text{(2.13)} \quad \begin{aligned} &\Pr\{\text{nonnull}|z\}/\Pr\{\text{null}|z\} \\ &= (1 - \text{fdr}(z))/\text{fdr}(z) \\ &= p_1 f_1(z)/p_0 f_0(z) \\ &\geq 0.8/0.2 = 4. \end{aligned}$$

Since (2.2) implies $p_1/p_0 \leq 1/9$, (2.13) corresponds to requiring Bayes factor

$$\text{(2.14)} \qquad f_1(z)/f_0(z) \geq 36$$

in favor of nonnull in order to declare significance.

Factor (2.14) requires much stronger evidence against the null hypothesis than in standard one-at-a-time testing, where the critical threshold lies somewhere near 3 (Efron and Gous, 2001). The fdr 0.20 threshold corresponds to $q$-values in (2.5) between 0.05 and 0.15 for moderate choices of $\gamma$; such $q$-value thresholds can be interpreted as providing conservative Bayes factors for Fdr testing.

Model (2.1) ignores the fact that investigators usually begin with hot prospects in mind, genes that have high prior probability of being interesting. Suppose $p_0(i)$ is the prior probability that gene $i$ is null, and define $p_0$ as the average of $p_0(i)$ over all $N$ genes. Then Bayes' theorem yields this expression for $\text{fdr}_i(z) = \Pr\{\text{gene}_i\ \text{null}|z_i = z\}$:

$$\text{(2.15)} \quad \begin{aligned} \text{fdr}_i(z) &= \text{fdr}(z) \frac{r_i}{1 - (1 - r_i)\text{fdr}(z)}, \\ &\left[ r_i = \frac{p_0(i)}{1 - p_0(i)} \Big/ \frac{p_0}{1 - p_0} \right], \end{aligned}$$

where $\text{fdr}(z) = p_0 f_0(z)/f(z)$ as before. So for a hot prospect having $p_0(i) = 0.50$ rather than $p_0 = 0.90$, (2.15) changes an uninteresting result like $\text{fdr}(z_i) = 0.40$ into $\text{fdr}_i(z_i) = 0.069$.

Wonderfully neat and exact results like the Benjamini–Hochberg Fdr control rule exert a powerful influence on statistical theory, sometimes more than is good for applied work. Much of the microarray statistics literature seems to me to be overly concerned with exact properties borrowed from classical test theory, at the expense of ignoring the complications of large-scale testing. Neatness and exactness are mostly missing in what follows as I examine an empirical Bayes approach to the application of two-groups/Fdr ideas to situations like those in Figure 1.

## 3. EMPIRICAL BAYES METHODS

In practice, the difference between Bayesian and frequentist statisticians is their self-confidence in assigning prior distributions to complicated probability models. Large-scale testing problems certainly look complicated enough, but this is deceptive; their massively parallel structure, with thousands of *similar* situations each providing information, allows an appropriate prior distribution to be estimated from the data without upsetting even timid frequentists like myself. This is the *empirical Bayes* approach of Robbins and Stein, 50 years old but coming into its own in the microarray era; see Efron (2003).

Consider estimating the local false discovery rate fdr$(z) = p_0 f_0(z)/f(z)$, (2.7). I will begin with a "good" case, like the prostate data of Example A in Section 1, where it is easy to believe in the theoretical null distribution (1.1),

$$(3.1) \qquad f_0(z) = \varphi(z) \equiv \frac{1}{\sqrt{2\pi}} e^{-(1/2)z^2}.$$

The $z$-values in Example A were obtained by transforming the usual two-sample $t$ statistic "$t_i$" comparing cancer and normal patients' expression levels for gene $i$, to a standard normal scale via

$$(3.2) \qquad z_i = \Phi^{-1}(F_{100}(t_i));$$

here $\Phi$ and $F_{100}$ are the cdf's of standard normal and $t_{100}$ distributions. If we had only gene $i$'s data to test, classic theory would tell us to compare $z_i$ with $f_0(z) = \varphi(z)$ as in (3.1).

For the moment I will take $p_0$, the prior probability of a gene being null, as known. Section 4 discusses $p_0$'s estimation, but in fact its exact value does not make much difference to Fdr$(z)$ or fdr$(z)$, (2.3) or (2.7), if $p_0$ is near 1 as in (2.2). Benjamini and Hochberg (1995) take $p_0 = 1$, providing an upper bound for Fdr$(z)$.

This leaves us with only the denominator $f(z)$ to estimate in (2.7). By definition (2.6), $f(z)$ is the marginal density of all $N$ $z_i$'s, so we can use all the data to estimate $f(z)$. The algorithm *locfdr*, an R function available from the CRAN library, does this by means of standard Poisson GLM software (Efron, 2005). Suppose the $z$-values have been binned, giving bin counts

$$(3.3) \quad y_k = \#\{z_i \text{ in bin } k\}, \quad k = 1, 2, \ldots, K.$$

The prostate data histogram in panel A of Figure 1 has $K = 49$ bins of width $\Delta = 0.2$.

We take the $y_k$ to be independent Poisson counts,

$$(3.4) \qquad y_k \stackrel{\text{ind}}{\sim} P_0(\nu_k), \quad k = 1, 2, \ldots, K,$$

with the unknown $\nu_k$ proportional to density $f(z)$ at midpoint "$x_k$" of the $k$th bin, approximately

$$(3.5) \qquad \nu_k = N\Delta f(x_k).$$

Modeling $\log(\nu_k)$ as a $p$th-degree polynomial function of $x_k$ makes (3.4)–(3.5) a standard Poisson general linear model (GLM). The choice $p = 7$ used in Figure 3 amounts to estimating $f(z)$ by maximum likelihood within the seven-parameter exponential family

$$(3.6) \qquad f(z) = \exp\left\{\sum_{j=0}^{7} \beta_j z^j\right\}.$$

Notice that $p = 2$ would make $f(z)$ normal; the extra parameters in (3.6) allow flexibility in fitting the tails of $f(z)$. Here we are employing *Lindsey's method*; see Efron and Tibshirani (1996). Despite its unorthodox look, it is no more than a convenient way to obtain maximum likelihood estimates in multiparameter families like (3.6).

The heavy curve in Figure 3 is an estimate of the local false discovery rate for the prostate data,

$$(3.7) \qquad \widehat{\text{fdr}}(z) = p_0 f_0(z)/\widehat{f}(z),$$

with $\widehat{f}(z)$ constructed as above, $f_0(z) = \varphi(z)$ as in (3.1), and $p_0 = 0.93$, as estimated in Section 4; $\widehat{\text{fdr}}(z)$ is near 1 for $|z| \le 2$, decreasing to interesting levels for $|z| > 3$. Fifty-one of the 6033 genes have $\widehat{\text{fdr}}(z_i) \le 0.2$, 26 on the right and 25 on the left, and these could be reported back to the investigators as likely nonnull candidates. [The standard Benjamini–Hochberg procedure, (2.5) with $q = 0.1$, reports 60 nonnull genes, 28 on the right and 32 on the left.]

At this point the reader might notice an anomaly: if $p_0 = 0.93$ of the $N$ genes are null, then about $(1 - p_0) \cdot 6033 = 422$ should be nonnull, but only 51 are reported. The trouble is that most of the nonnull genes are located in regions of the $z$ axis where $\widehat{\text{fdr}}(z_i)$ exceeds 0.5, and these cannot be reported

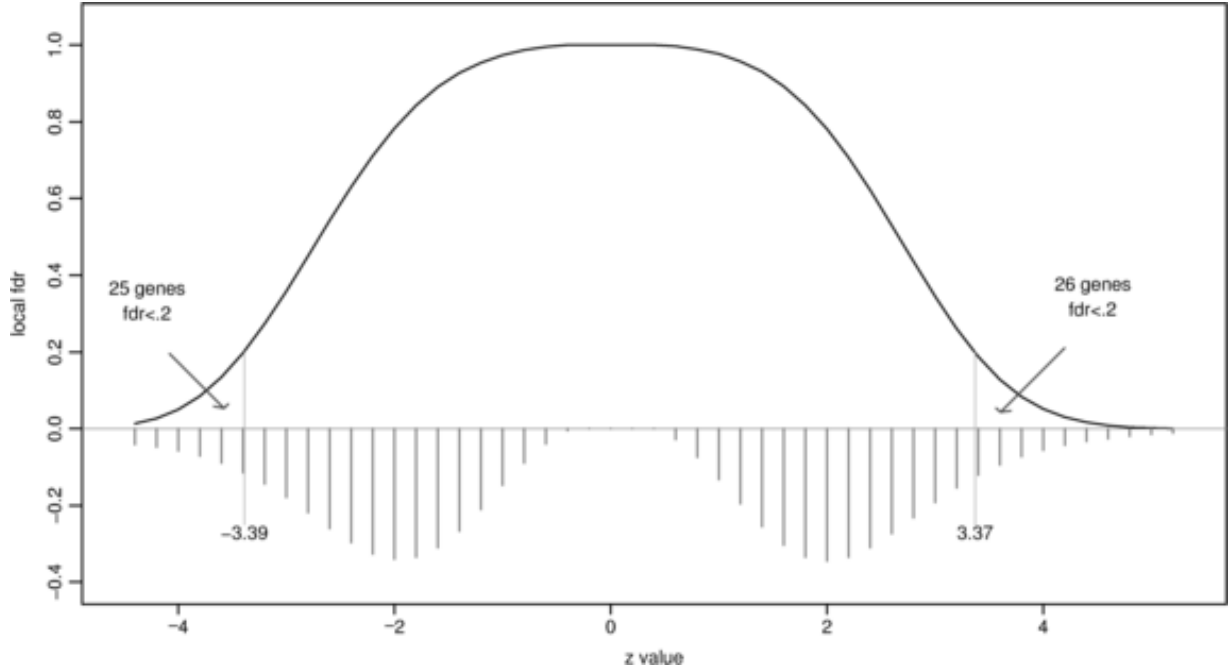

FIG. 3. *Heavy curve is estimated local false discovery rate* $\widehat{fdr}(z)$ *for prostate data. Fifty-one genes, 26 on the right and 25 on the left, have* $\widehat{fdr}(z_i) < 0.20$. *Vertical bars estimate histogram of the nonnull counts (plotted negatively, divided by 50). Most of the nonnull genes will not be reported.*

without also reporting a bevy of null cases. In other words, the prostate study is underpowered.

The vertical bars in Figure 3 are estimates of the *nonnull counts*, the histogram we would see if only the nonnull genes provided $z$-values. In terms of (3.3), (3.7), the nonnull counts "$y_k^{(1)}$" are

$$(3.8) \qquad y_k^{(1)} = [1 - \widehat{\mathrm{fdr}}_k] y_k,$$

where $\widehat{\mathrm{fdr}}_k = \widehat{\mathrm{fdr}}(x_k)$, the estimated fdr value at the center of bin $k$. Since $1 - \widehat{\mathrm{fdr}}_k$ approximates the nonnull probability for a gene in bin $k$, formula (3.8) is an obvious estimate for the expected number of nonnulls.

Power diagnostics are obtained from comparisons of $\widehat{\mathrm{fdr}}(z)$ with the nonnull histogram. High power would be indicated if $\widehat{\mathrm{fdr}}_k$ was small where $y_k^{(1)}$ was large. That obviously is not the case in Figure 3. A simple power diagnostic is

$$(3.9) \qquad \widehat{E\,\mathrm{fdr}}^{(1)} = \sum_{k=1}^{K} y_k^{(1)} \widehat{\mathrm{fdr}}_k \Big/ \sum_{k=1}^{K} \widehat{y}_k^{(1)},$$

the expected nonnull fdr. We want $\widehat{E\,\mathrm{fdr}}^{(1)}$ to be small, perhaps near 0.2, so that a typical nonnull gene will show up on a list of likely prospects. The prostate data has $\widehat{E\,\mathrm{fdr}}^{(1)} = 0.68$, indicating low power. If the whole study were rerun, we could expect a different list of 50 likely nonnull genes, barely overlapping with the first list. Section 3 of Efron (2006) discusses power calculations for microarray studies, presenting more elaborate power diagnostics.

Stripped of technicalities, the idea underlying false discovery rates is appealingly simple, and in fact does not depend on the literal validity of the two-groups model (2.1). Consider the bin $z_i \in [3.1, 3.3]$ in the prostate data histogram; 17 of the 6033 genes fall into this bin, compared to expected number $2.68 = p_0 N \Delta \varphi(3.2)$ of null genes, giving

$$(3.10) \qquad \overline{\mathrm{fdr}} = 2.68/17 = 0.16$$

as an estimated false discovery rate. (The smoothed estimate in Figure 3 is $\widehat{\mathrm{fdr}} = 0.24$.) The implication is that only about one-sixth of the 17 are null genes. This conclusion can be sharpened, as in Lehmann and Romano (2005), but (3.10) catches the main idea.

Notice that we do not need all the null genes to have the *same* density $f_0(z)$; it is enough to assume that the *average* null density is $f_0(z)$, $\varphi(z)$ in this case, in order to calculate the numerator 2.68.

TABLE 1
*Boldface, standard errors of $\log \widehat{\mathrm{fdr}}(z)$, (local fdr), and $\log \widehat{\mathrm{Fdr}}(z)$, (tail-area), 250 replications of model (3.11), $N = 1500$. Parentheses, average from formula (5.9), Efron (2006); fdr is true value (2.7). Empirical Null results explained in Section 4*

|  |  | **Theoretical null** | | | **Empirical null** | | |
| --- | --- | --- | --- | --- | --- | --- | --- |
| $z$ | fdr | local | (formula) | tail | local | (formula) | tail |
| 1.5 | 0.88 | **0.05** | (0.05) | **0.05** | **0.04** | (0.04) | **0.10** |
| 2.0 | 0.69 | **0.08** | (0.09) | **0.05** | **0.09** | (0.10) | **0.15** |
| 2.5 | 0.38 | **0.09** | (0.10) | **0.05** | **0.16** | (0.16) | **0.23** |
| 3.0 | 0.12 | **0.08** | (0.10) | **0.06** | **0.25** | (0.25) | **0.32** |
| 3.5 | 0.03 | **0.10** | (0.13) | **0.07** | **0.38** | (0.38) | **0.42** |
| 4.0 | 0.005 | **0.11** | (0.15) | **0.10** | **0.50** | (0.51) | **0.52** |

(This is an advantage of false discovery rate methods, which only control *rates*, not *individual probabilities*.) The nonnull density $f_1(z)$ in (2.1) plays no role at all since the denominator 17 is an observed quantity. *Exchangeability* is the key assumption in interpreting (3.10): we expect about 1/6 of the 17 genes to be null, and assign posterior null probability 1/6 to all 17. Nonexchangeability, in the form of differing prior information among the 17, can be incorporated as in (2.15).

Density estimation has a reputation for difficulty, well-deserved in general situations. However, there are good theoretical reasons, presented in Section 6 of Efron (2005), for believing that mixtures of $z$-values are quite smooth, and that (3.7) will efficiently estimate fdr($z$). Independence of the $z_i$'s is *not* required, only that $\widehat{f}(z)$ is a reasonably close estimate of $f(z)$.

Table 1 reports on a small simulation study in which

$$(3.11) \quad z_i \stackrel{\text{ind}}{\sim} N(\mu_i, 1) \begin{cases} \mu_i = 0, \\ \quad \text{with probability } 0.9, \\ \mu_i \sim N(3, 1), \\ \quad \text{with probability } 0.1, \end{cases}$$

for $i = 1, 2, \ldots, N = 1500$. The table shows standard deviations for $\log(\widehat{\mathrm{fdr}}(z))$, (3.7), from 250 simulations of (3.11), and also using a delta-method formula derived in Section 5 of Efron (2006), incorporated in the *locfdr* algorithm. Rather than (3.6), $f(z)$ was modeled by a seven-parameter natural spline basis, *locfdr*'s default, though this gave nearly the same results as (3.6). Also shown are standard deviations for the corresponding tail-area quantity $\log(\widehat{\mathrm{Fdr}}(z))$ obtained by substituting $\widehat{F}(z) = \int_{-\infty}^{z} \widehat{f}(z')\,dz'$ in (2.3). [This is a little less variable than using $\bar{F}(z)$, (2.4).]

The "Theoretical Null" side of the table shows that $\widehat{\mathrm{fdr}}(z)$ is more variable than $\widehat{\mathrm{Fdr}}(z)$, but both are more than accurate enough for practical use. At $z = 3$, for example, $\widehat{\mathrm{fdr}}(z)$ only errs by about 8%, yielding $\widehat{\mathrm{fdr}}(z) \doteq 0.12 \pm 0.01$. Standard errors are roughly proportional to $N^{-1/2}$, so even reducing $N$ to 250 gives $\widehat{\mathrm{fdr}}(3) \doteq 0.12 \pm .025$, and similarly for other values of $z$, accurate enough to make pictures like Figure 3 believable.

Empirical Bayes is a bipolar methodology, with alternating episodes of frequentist and Bayesian activity. Frequentists may prefer $\widehat{\mathrm{Fdr}}$ [or $\overline{Fdr}$, (2.5)] to $\widehat{\mathrm{fdr}}$ because of connections with classical tail-area hypothesis testing, or because cdf's are more straightforward to estimate than densities, while Bayesians prefer $\widehat{\mathrm{fdr}}$ for its more apt a posteriori interpretation. Both, though, combine the Bayesian two-groups model with frequentist estimation methods, and deliver the same basic information.

A variety of local fdr estimation methods have been suggested, using parametric, semiparametric, nonparametric and Bayes methods: Pan et al. (2003), Pounds and Morris (2003), Allison et al. (2002), Heller and Qing (2003), Broberg (2005), Aubert et al. (2004), Liao et al. (2004) and Do et al. (2005), all performing reasonably well. The Poisson GLM methodology of *locfdr* has the advantage of easy implementation with familiar software, and a closed-form error analysis.

Estimation efficiency becomes a more serious problem on the "Empirical Null" side of Table 1, where we can no longer trust the theoretical null $f_0(z) \sim N(0, 1)$. This is the subject of Section 4.

## 4. THE EMPIRICAL NULL DISTRIBUTION

We have been assuming that $f_0(z)$, the null density in (2.1), is known on theoretical grounds, as in (3.1). This leads to false discovery estimates such as $\widehat{\mathrm{fdr}}(z) = p_0 f_0(z) / \widehat{f}(z)$ and $\widehat{\mathrm{Fdr}}(z) = p_0 F_0(z) / \widehat{F}(z)$, where only denominators need be estimated. Most applications of Benjamini and Hochberg's control algorithm (2.5) make the same assumption (sometimes augmented with permutation calculations, which usually produce only minor corrections to the theoretical null, as discussed in Section 5). Use of the theoretical null is mandatory in classic one-at-a-time testing, where theory provides the only information available for null behavior. But things change in large-scale simultaneous testing situations: serious defects in the theoretical null may become obvious, while empirical Bayes methods can provide more realistic null distributions.

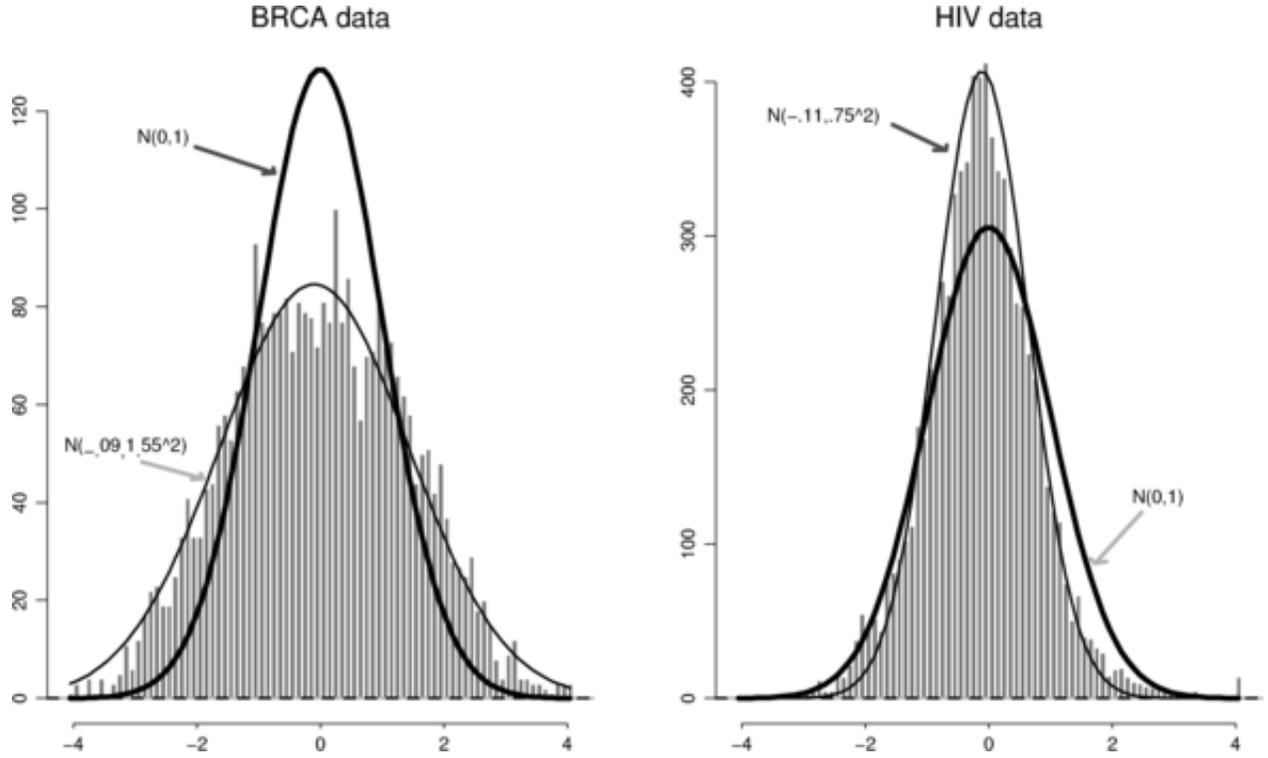

FIG. 4. *z-values from two microarray studies. BRCA data (Hedenfalk et al., 2001), comparing seven breast cancer patients having BRCA1 mutation to eight with BRCA2 mutation $N = 3226$ genes. HIV data (van't Wout et al., 2003) comparing four HIV+ males with four HIV− males, $N = 7680$ genes. Theoretical $N(0,1)$ null, heavy curve is too narrow for BRCA data, too wide for HIV data. Light curves are empirical nulls: normal densities fit to the central histogram counts.*

Figure 4 shows $z$-value histograms for two additional microarray studies, described more fully in Efron (2006). These are of the same form as the prostate data: $n$ subjects in two disease categories provide expression levels for $N$ genes; two-sample $t$-statistics $t_i$ comparing the categories are computed for each gene, and then transformed to $z$-values $z_i = \Phi^{-1}(F_{n-2}(t_i))$, as in (3.2). Unlike panel A of Figure 1, however, neither histogram obeys the theoretical $N(0,1)$ null near $z=0$. The BRCA data has a much wider central peak, while the HIV peak is too narrow. The lighter curves in Figure 4 are *empirical null* estimates (Efron, 2004), normal curves fit to the central peak of the $z$-value histograms. The idea here is simple enough: we make the "zero assumption,"

ZERO ASSUMPTION.

(4.1) Most of the $z$-values near 0 come from null genes,

(discussed further below), generalize the $N(0,1)$ theoretical null to $N(\delta_0, \sigma_0^2)$, and estimate $(\delta_0, \sigma_0^2)$ from the histogram counts near $z=0$. *Locfdr* uses two different estimation methods, analytical and geometric, described next.

Figure 5 shows the geometric method in action on the HIV data. The heavy solid curve is $\log \widehat{f}(z)$, fit from (3.6) using Lindsey's method, as described in Efron and Tibshirani (1996). The two-groups model and the zero assumption suggest that if $f_0$ is normal, $f(z)$ should be well-approximated near $z=0$ by $p_0 \varphi_{\delta_0, \sigma_0}(z)$, with

$$(4.2) \quad \varphi_{\delta_0, \sigma_0}(z) \equiv (2\pi\sigma_0^2)^{-1/2} \exp\left\{-\frac{1}{2}\left(\frac{z-\delta_0}{\sigma_0}\right)^2\right\},$$

making $\log f(z)$ approximately quadratic,

$$(4.3) \quad \log f(z) \doteq \log p_0 - \frac{1}{2}\left\{\frac{\delta_0^2}{\sigma_0^2} + \log(2\pi\sigma_0^2)\right\} + \frac{\delta_0}{\sigma_0^2} z - \frac{1}{2\sigma_0^2} z^2.$$

The beaded curve shows the best quadratic approximation to $\log \widehat{f}(z)$ near 0. Matching its coefficients

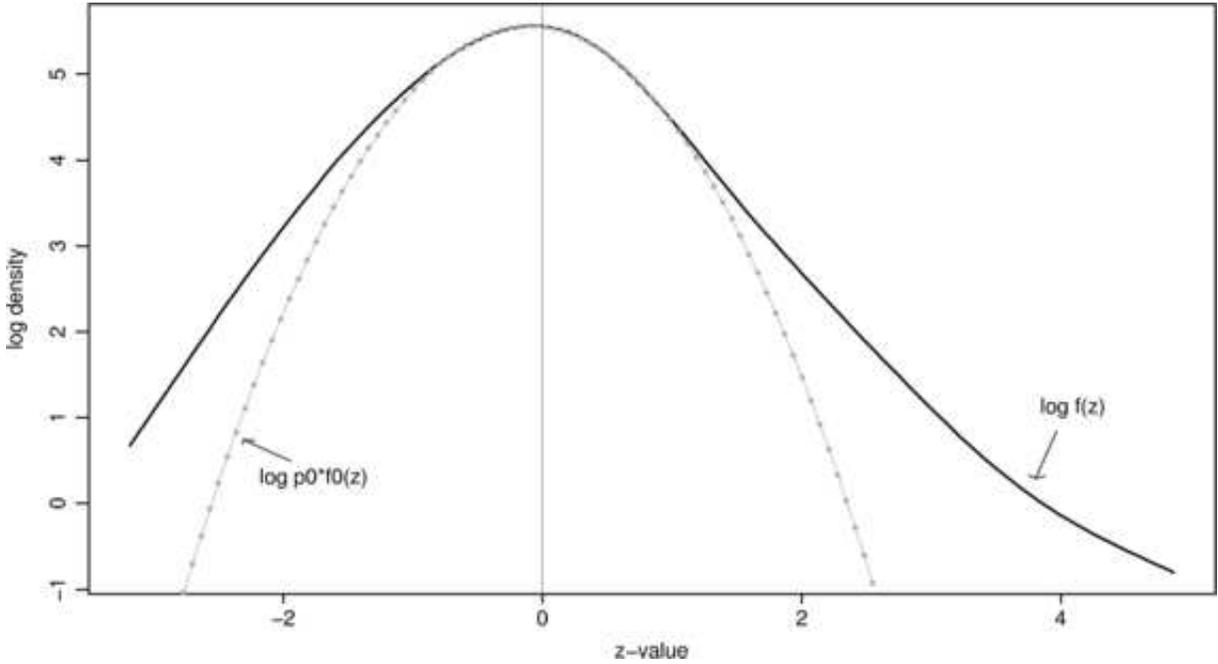

Fig. 5. *Geometric estimate of null proportion $p_0$ and empirical null mean and standard deviation $(\delta_0, \sigma_0)$ for the HIV data. Heavy curve is $\log \widehat{f}(z)$, estimated as in (3.3)–(3.6); beaded curve is best quadratic approximation to $\log \widehat{f}(z)$ near $z = 0$.*

$(\widehat{\beta}_0, \widehat{\beta}_1, \widehat{\beta}_2)$ to (4.3) yields estimates $(\widehat{\delta}_0, \widehat{\sigma}_0, \widehat{p}_0)$, for instance, $\widehat{\sigma}_0 = (2\widehat{\beta}_2)^{-1/2}$,

$$\widehat{\delta}_0 = -0.107,$$
(4.4) $$\widehat{\sigma}_0 = 0.753,$$
$$\widehat{p}_0 = 0.931,$$

for the HIV data. Trying the same method with the theoretical null, that is, taking $(\delta_0, \sigma_0) = (0, 1)$ in (4.3), gives a very poor fit, and $\widehat{p}_0$ equals the impossible value 1.20.

The analytic method makes more explicit use of the zero assumption, stipulating that the nonnull density $f_1(z)$ in the two-groups model (2.1) is supported outside some given interval $[a, b]$ containing zero (actually chosen by preliminary calculations). Let $N_0$ be the number of $z_i$ in $[a, b]$, and define

$$P_0(\delta_0, \sigma_0) = \Phi\left(\frac{b - \delta_0}{\sigma_0}\right) - \Phi\left(\frac{a - \delta_0}{\sigma_0}\right) \quad \text{and}$$
(4.5)
$$\theta = p_0 P_0.$$

Then the likelihood function for $\mathbf{z}_0$, the vector of $N_0$ $z$-values in $[a, b]$, is

(4.6) $$f_{\delta_0, \sigma_0, p_0}(\mathbf{z}_0) = [\theta^{N_0}(1-\theta)^{N-N_0}]$$
$$\cdot \left[\prod_{z_i \in \mathbf{z}_0} \frac{\varphi_{\delta_0, \sigma_0}(z_i)}{P_0(\delta_0, \sigma_0)}\right].$$

This is the product of two exponential family likelihoods, which is numerically easy to solve for the maximum likelihood estimates $(\widehat{\delta}_0, \widehat{\sigma}_0, \widehat{p}_0)$, equaling $(-0.120, 0.787, 0.956)$ for the HIV data.

Both methods are implemented in *locfdr*. The analytic method is somewhat more stable but can be more biased than geometric fitting. Efron (2004) shows that geometric fitting gives nearly unbiased estimates of $\delta_0$ and $\sigma_0$ for $p_0 \geq 0.90$. Table 2 shows how the two methods fared in the simulation study of Table 1.

A healthy literature has sprung up on the estimation of $p_0$, as in Pawitan et al. (2005) and Langlass et al. (2005), all of which assumes the validity of the theoretical null. The zero assumption plays a central role in this literature [which mostly works with two-sided $p$-values rather than $z$-values, e.g., $p_i = 2(1 - F_{100}(|t_i|))$ in (3.2), making the "zero region" occur near $p = 1$]. *The two-groups model is unidentifiable if $f_0$ is unspecified in* (2.1), since we can redefine $f_0$ as $f_0 + cf_1$, and $p_1$ as $p_1 - cp_0$ for any $c \leq p_1/p_0$. With $p_1$ small, (2.2), and $f_1$ supposed to yield $z_i$'s far from 0 for the most part, the zero assumption is a reasonable way to impose identifiability on the two-groups model. Section 6 considers the meaning of the null density more carefully, among other things explaining the upward bias of $\hat{p}_0$ seen in Table 2.

Table 2
*Comparison of estimates $(\widehat{\delta}_0, \widehat{\sigma}_0, \widehat{p}_0)$, simulation study of Table 1. "Formula" is average from delta-method standard deviation formulas, Section 5 in Efron (2006), as implemented in locfdr*

| | **Geometric** | | | **Analytic** | | |
|---|---|---|---|---|---|---|
| | **mean** | **stdev** | **(formula)** | **mean** | **stdev** | **(formula)** |
| $\widehat{\delta}_0$: | 0.02 | **0.056** | (0.062) | 0.04 | **0.031** | (0.032) |
| $\widehat{\sigma}_0$: | 1.02 | **0.029** | (0.033) | 1.04 | **0.031** | (0.031) |
| $\widehat{p}_0$: | 0.92 | **0.013** | (0.015) | 0.93 | **0.009** | (0.011) |

The empirical null is an expensive luxury from the point of view of estimation efficiency. Comparing the right-hand side of Table 1 with the left reveals factors of 2 or 3 increase in standard error relative to the theoretical null, near the crucial point where $\text{fdr}(z) = 0.2$. Section 4 of Efron (2005) pins the increased variability entirely on the estimation of $(\delta_0, \sigma_0)$; even knowing the true values of $p_0$ and $f(z)$ would reduce the standard error of $\log \widehat{\text{fdr}}(z)$ by less than 1%. (Using tail-area Fdr's rather than local fdr's does not help—here the local version is less variable.)

The reason for considering empirical nulls is that the theoretical $N(0, 1)$ null does not seem to fit the data in situations like Figure 4. For the BRCA data we can see that the histogram is overdispersed compared to $N(0, 1)$ around $z = 0$; the implication is that there will be more null counts far from zero than the theoretical null predicts, making $N(0, 1)$ false discovery rate calculations like (3.10) too optimistic. The opposite happens with the HIV data.

There is a lot at stake here for both Bayesians and frequentists. Table 3 shows the number of gene discoveries identified by the standard Benjamini–Hochberg two-sided Fdr procedure, $q = 0.10$ in (2.5). The HIV results are much more dramatic using the empirical null $f_0(z) \sim N(-0.11, 0.75^2)$ and in fact we will see in the next section that $\sigma_0 = 0.75$ is quite believable in this case. The BRCA data has been used in the microarray literature to compare analysis techniques, under the presumption that better techniques will produce more discoveries; recently, for instance, in Storey et al. (2005) and Pawitan et al. (2005). Table 3 suggests caution in this interpretation, where using the empirical null negates any discoveries at all.

The $z$-values in panel C of Figure 1, proteomics data, were calculated from standard Cox likelihood

TABLE 3
*Number of genes identified as true discoveries by two-sided Benjamini–Hochberg procedure, 0.10 control level*

|  | **Theoretical null** | **Empirical null** |
|---|---|---|
| BRCA data: | 107 | 0 |
| HIV data: | 22 | 180 |

Empirical null densities as in Figure 4.

tests that should yield $N(0, 1)$ null results asymptotically. A $N(-0.02, 1.29^2)$ empirical null was obtained from the analytic method, resulting in only one peak with $\widehat{\text{fdr}} < 0.2$; using the theoretical null gave six such peaks.

In panel B of Figure 1, the $z$-values were obtained from familiar binomial calculations, each $z_i$ being calculated as

$$(4.7) \quad z = (\hat{p}_{\text{ad}} - \hat{p}_{\text{dis}} - \Delta) \\ \cdot \left( \frac{\hat{p}_{\text{ad}}(1 - \hat{p}_{\text{ad}})}{n_{\text{ad}}} + \frac{\hat{p}_{\text{dis}}(1 - \hat{p}_{\text{dis}})}{n_{\text{dis}}} \right)^{-1/2},$$

where $n_{\text{ad}}$ was the number of advantaged students in the high school, $\hat{p}_{\text{ad}}$ the proportion passing the test, and likewise $n_{\text{dis}}$ and $\hat{p}_{\text{dis}}$ for the disadvantaged students; $\Delta = 0.192$ was the overall difference, median $(\hat{p}_{\text{ad}})$ − median $(\hat{p}_{\text{dis}})$. Here the empirical null standard deviation $\hat{\sigma}_0$ equals 1.52, half again bigger than the theoretical standard deviation we would use if we had only one school's data. An empirical null fdr analysis yielded 75 schools with $\widehat{\text{fdr}} < 0.20$, 30 on the left and 45 on the right. Example B is discussed a bit further in the next two sections, where its use in the two-groups model is questioned.

My point here is not that the empirical null is always the correct choice. The opposite advice, always use the theoretical null, has been inculcated by a century of classic one-case-at-a-time testing to the point where it is almost subliminal, but it exposes the statistician to obvious criticism in situations like the BRCA and HIV data. Large-scale simultaneous testing produces mass information of a Bayesian nature that impinges on individual decisions. The two-groups model helps bring this information to bear, after one decides on the proper choice of $f_0$ in (2.1). Section 5 discusses this choice, in the form of a list of reasons why the theoretical null, and its close friend the permutation null, might go astray.

## 5. THEORETICAL, PERMUTATION AND EMPIRICAL NULL DISTRIBUTIONS

Like most statisticians, I have spent my professional life happily testing hypotheses against theoretical null distributions. It came as somewhat of a shock then, when pictures like Figure 4 suggested that the theoretical null might be more theoretical than I had supposed. Once suspicious, it becomes easy to think of reasons why $f_0(z)$, the crucial element in the two-groups model (2.1), might not obey classical guidelines. This section presents four reasons why the theoretical null might fail, and also gives me a chance to say something about the strengths and weaknesses of permutation null distributions.

REASON 1 (*Failed mathematical assumptions*). The usual derivation of the null hypothesis distribution for a two-sample $t$-statistic assumes independent and identically distributed (i.i.d.) normal components. For the BRCA data of Figure 4, direct inspection of the 3226 by 15 matrix "$X$" of expression values reveals markedly nonnormal components, skewed to the right (even after the columns of $X$ have been standardized to mean 0 and standard deviation 1, as in all my examples here). Is this causing the failure of the $N(0,1)$ theoretical null?

Permutation techniques offer quick relief from such concerns. The columns of $X$ are randomly permuted, giving a matrix $X^*$ with corresponding $t$-values $t_i^*$ and $z$-values $z_i^* = \Phi^{-1}(F_{n-2}(t_i^*))$. This is done some large number of times, perhaps 100, and the empirical distribution of the $100 \cdot N$ $z_i^*$'s used as a *permutation null*. The well-known SAM algorithm (Tusher, Tibshirani and Chu, 2001) effectively employs the permutation null cdf in the numerator of the Fdr formula (2.3).

Applied to the BRCA matrix, the permutation null came out nearly $N(0,1)$ (as did simply simulating the entries of $X^*$ by independent draws from all $3226 \cdot 15$ entries of $X$), so nonnormal distributions were not the cause of BRCA's overwide histogram. In practice the permutation null usually approximates the theoretical null closely, as a long history of research on the permutation $t$-test demonstrated; see Section 5.9 of Lehmann and Romano (2005).

REASON 2 (*Unobserved covariates*). The BRCA study is observational rather than experimental—the 15 women were *observed* to be BRCA1 or BRCA2, not *assigned*, and likewise with the HIV and prostate studies. There are likely to be covariates—age, race, general health—that affect the microarray expression levels differently for different genes. If these were known to us, they could be factored out using a separate linear model on each gene's data, providing a new and improved $z_i$ obtained from the "Treatment" coefficient in the model. This would reduce the spread of the $z$-value histogram, perhaps even restoring the $N(0,1)$ theoretical null for the BRCA data.

Unobserved covariates act to broaden the null distribution $f_0(z)$. They also broaden the nonnull distribution $f_1(z)$ in (2.1), and the mixture density $f(z)$, but this does not correct fdr estimates like (3.10), where the numerator, which depends entirely on $f_0$, is the only estimated quantity. Section 4 of Efron (2004) provides an analysis of a simplified model with unobserved covariates. Permutation techniques cannot recognize unobserved covariates, as the model demonstrates.

REASON 3 (*Correlation across arrays*). False discovery rate methodology does not require independence among the test statistics $z_i$. However, the theoretical null distribution does require independence of the expression values used to calculate each $z_i$; in terms of the elements $x_{ij}$ of the expression matrix $X$, for gene $i$ we need independence among $x_{i1}, x_{i2}, \ldots, x_{in}$ in order to validate (1.1).

Experimental difficulties can undercut across-microarray independence, while remaining undetectable in a permutation analysis. This happened in both studies of Figure 4 (Efron, 2004, 2006). The BRCA data showed strong positive correlations among the first four BRCA2 arrays, and also among the last four. This reduces the effective degrees of freedom for each $t$-statistic below the nominal 13, making $t_i$ and $z_i = \Phi^{-1}(F_{13}(t_i))$ overdispersed.

REASON 4 (*Correlation across genes*). Benjamini and Hochberg's 1995 paper verified Fdr control for rule (2.5) under the assumption of independence among the $N$ $z$-values (relaxed a little in Benjamini and Yekutieli, 2001). This seems fatal for microarray applications since we expect genes to be correlated in their actions. A great virtue of the empirical Bayes/two-groups approach is that independence is not necessary; with $\widehat{\text{Fdr}}(z) = p_0 F_0(z)/\widehat{F}(z)$, for instance, $\widehat{\text{Fdr}}(z)$ can provide a reasonable estimate of $\Pr\{\text{null}|Z \leq z\}$ as long as $\widehat{F}(z)$ is roughly unbiased for $F(z)$—in formal terms requiring consistency but not independence—and likewise for the local version $\widehat{\mathrm{fdr}}(z) = p_0 f_0(z)/\widehat{f}(z)$, (3.7).

There is, however, a black cloud inside the silver lining: the assumption that the null density $f_0(z)$ is known to the statistician. The empirical null estimation methods of Section 4 do not require $z$-value independence, and so disperse the black cloud, at the expense of increased variability in fdr estimates. Do we really need to use an empirical null? Efron (2007) discusses the following somewhat disconcerting result: even if the theoretical null distribution $z_i \sim N(0,1)$ holds exactly true for all null genes, Reasons 1–3 above not causing trouble, correlation among the $z_i$'s can make the overall null distribution effectively much wider or much narrower than $N(0,1)$.

Microarray data sets tend to have substantial $z$-value correlations. Consider the BRCA data: there are more than five million correlations $\rho_{ij}$ between pairs of gene $z$-values $z_i$ and $z_j$; by examining the row-wise correlations in the $X$ matrix we can estimate that the distribution of the $\rho_{ij}$'s has approximately mean 0 and variance $\alpha^2 = 0.153^2$,

$$(5.1) \qquad \rho \sim (0, \alpha^2).$$

(The zero mean is a consequence of standardizing the columns of $X$.) This is a lot of correlation—as much as if the BRCA genes occurred in 10 independent groups, but with common interclass correlation 0.50 for all genes within a group.

Section 3 of Efron (2006) shows that under assumptions (1.1)–(5.1), the ensemble of null-gene $z$-values will behave roughly as

$$(5.2) \qquad z_i \dot\sim N(0, \sigma_0^2)$$

with

$$(5.3) \qquad \sigma_0^2 = 1 + \sqrt{2} A, \quad A \sim (0, \alpha^2).$$

If the variable $A$ equaled $\alpha = 0.153$, for instance, giving $\sigma_0 = 1.10$, then the expected number of null counts below $z = -3$ would be about $p_0 N \Phi(-3/1.10)$ rather than $p_0 N \Phi(-3)$, more than twice as many. There is even more correlation in the HIV data, $\alpha \doteq 0.42$, enough so that a moderately negative value of $A$ could cause $\sigma_0 = 0.75$, as in Figure 4.

The random variable $A$ acts like an observable ancillary in the two-groups situation—observable because we can estimate $\sigma_0$ from the central counts of the $z$-value histogram, as in Section 4; $\widehat{\sigma}_0$ is essentially the half-width of the central peak.

Figure 6 is a cautionary story on the dangers of ignoring $\widehat{\sigma}_0$. A simulation model with

$$(5.4) \quad \begin{aligned} z_i &\sim N(0,1), \quad i = 1, 2, \ldots, 2700, \quad \text{and} \\ z_i &\sim N(2.5, 1.5), \quad i = 2701, \ldots, 3000, \end{aligned}$$

was run, in which the null $z_i$'s, the first 2700, were correlated to the same degree as in the BRCA data, $\alpha = 0.153$. For each of 1000 simulations of (5.4), a standard Benjamini–Hochberg Fdr analysis (2.5) (i.e., using the theoretical null for $F_0$) was run at control level $q = 0.10$, and used to identify a set of nonnull genes.

Each of the thousand points in Figure 6 is $(\widehat{\sigma}_0, \mathrm{Fdp})$, where $\widehat{\sigma}_0$ is half the distance between the 16th and 86th percentiles of the 3000 $z_i$'s, and Fdp is the "False discovery proportion," the proportion of identified genes that were actually null. Fdp averaged 0.091, close to the target value $q = 0.10$, but with a strong dependence on $\widehat{\sigma}_0$: the lowest 5% of $\widehat{\sigma}_0$'s corresponded to Fdp's averaging only 0.03, while the upper 5% average was 0.29, a factor of 9 difference.

The point here is not that the claimed $q$-value 0.10 is wrong, but that in any one simulation we may be able to see, from $\widehat{\sigma}_0$, that it is probably misleading. Using the empirical null counteracts this fallacy which, again, is not apparent from the permutation null. (Section 4 of Efron, 2007, discusses more elaborate permutation methods that do bear on Figure 6. See Qui et al., 2005, for a gloomier assessment of correlation effects in microarray analyses.)

What is causing the overdispersion in the Education data of panel B, (4.7)? Correlation across schools, Reason 44, seems ruled out by the nature of the sampling, leaving Reasons 2 and 3 as likely candidates; unobserved covariates are an obvious threat here, while within-school sampling dependences (Reason 3) are certainly possible. Fdr analysis yields eight times as many "significant" schools based on the theoretical null rather than $f_0 \sim N(-0.35, 1.51^2)$, but looks completely untrustworthy to me.

Sometimes the theoretical null distribution is fine, of course. The prostate data had $(\widehat{\delta}_0, \widehat{\sigma}_0) = (0.00, 1.06)$ according to the analytic method of (4.6), close enough to $(0,1)$ to make theoretical null calculations believable. However, there are lots of things that can go wrong with the theoretical null, and lots of data to check it with in large-scale testing situations, making it a matter of due diligence for the statistician to do such checking, even if only by visual inspection of the $z$-value histogram. All simultaneous testing procedures, not just false discovery rates, go wrong if the null distribution is misrepresented.

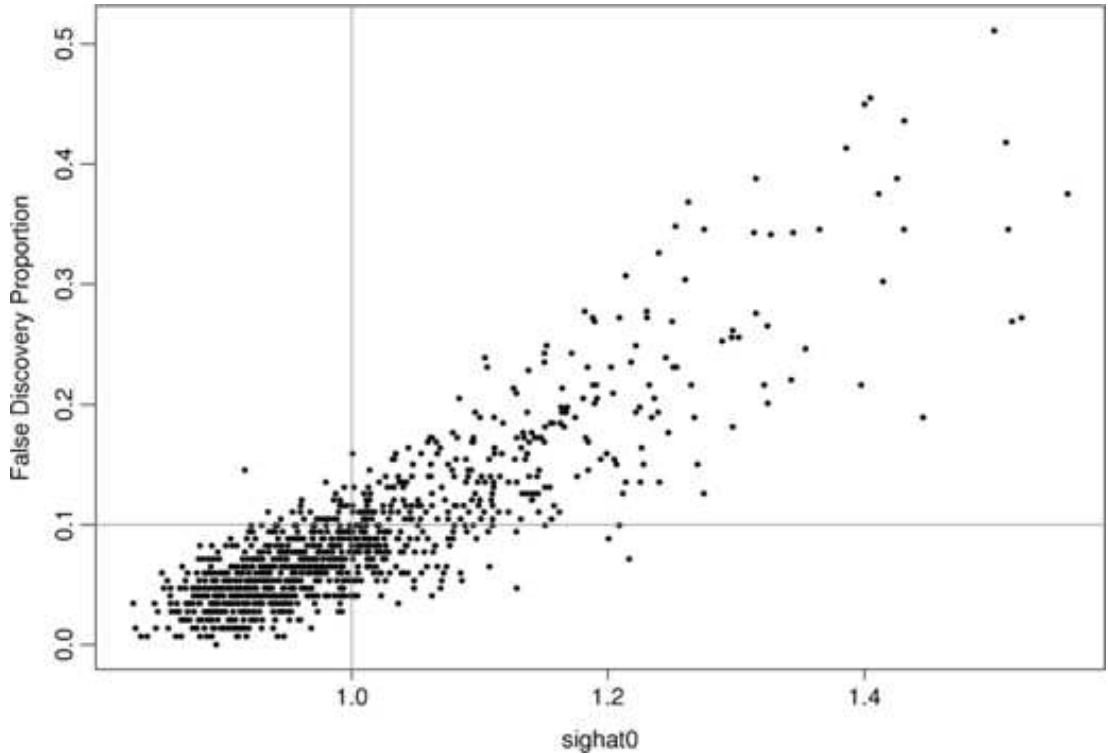

FIG. 6. *Benjamini–Hochberg Fdr control procedure (2.5), $q = 0.1$, run for 1000 simulations of correlated model (5.4); true false discovery proportion Fdp plotted versus half-width estimate $\widehat{\sigma}_0$. Overall Fdp averaged 0.091, close to q, but with a strong dependence on $\widehat{\sigma}_0$.*

## 6. A ONE-GROUP MODEL

Classical one-at-a-time hypothesis testing depends on having a unique null density $f_0(z)$, such as Student's $t$ distribution for the normal two-sample situation. The assumption of unique $f_0$ has been carried over into most of the microarray testing literature, including our definition (2.1) of the two-groups model.

Realistic examples of large-scale inference are apt to be less clearcut, with true effect sizes ranging continuously from zero or near zero to very large. Here we consider a "one-group" structural model that allows for a range of effects. We can still usefully apply fdr methods to data from one-group models; doing so helps clarify the choice between theoretical and empirical null hypotheses, and explicates the biases inherent in model (2.1). The discussion in this section, as in Section 2, will be mostly theoretical, involving probability models rather than collections of observed $z$-values.

Model (2.1) does not require knowing how the $z$-values were generated, a substantial practical advantage of the two-groups formulation. In contrast, one-group analysis begins with a specific Bayesian structural model. We assume that the $i$th case has an unobserved *true value* $\mu_i$ distributed according to some density $g(\mu)$, and that the observed $z_i$ is normally distributed around $\mu_i$,

(6.1) $\quad \mu \sim g(\cdot) \quad \text{and} \quad z|\mu \sim N(\mu, 1).$

The density $g(\mu)$ is allowed to have discrete atoms. It might have an atom at zero but this is not required, and in any case there is no a priori partition of $g(\mu)$ into null and nonnull components.

As an example, suppose $g(\mu)$ is a mixture of 90% $N(0, 0.5^2)$ and 10% $N(2.5, 0.5^2)$,

(6.2) $\quad g(\mu) = 0.9 \cdot \varphi_{0, 0.5}(\mu) + 0.1 \cdot \varphi_{2.5, 0.5}(\mu)$

in notation (4.2). The histogram in Figure 7 shows $N = 3000$ draws of $\mu_i$ from (6.2). I am thinking of this as a situation having a large proportion of uninteresting cases centered near, but not exactly at, zero, and a small proportion of interesting cases centered far to the right. We still want to use the observed $z_i$'s from (6.2) to flag cases that are likely to be interesting.

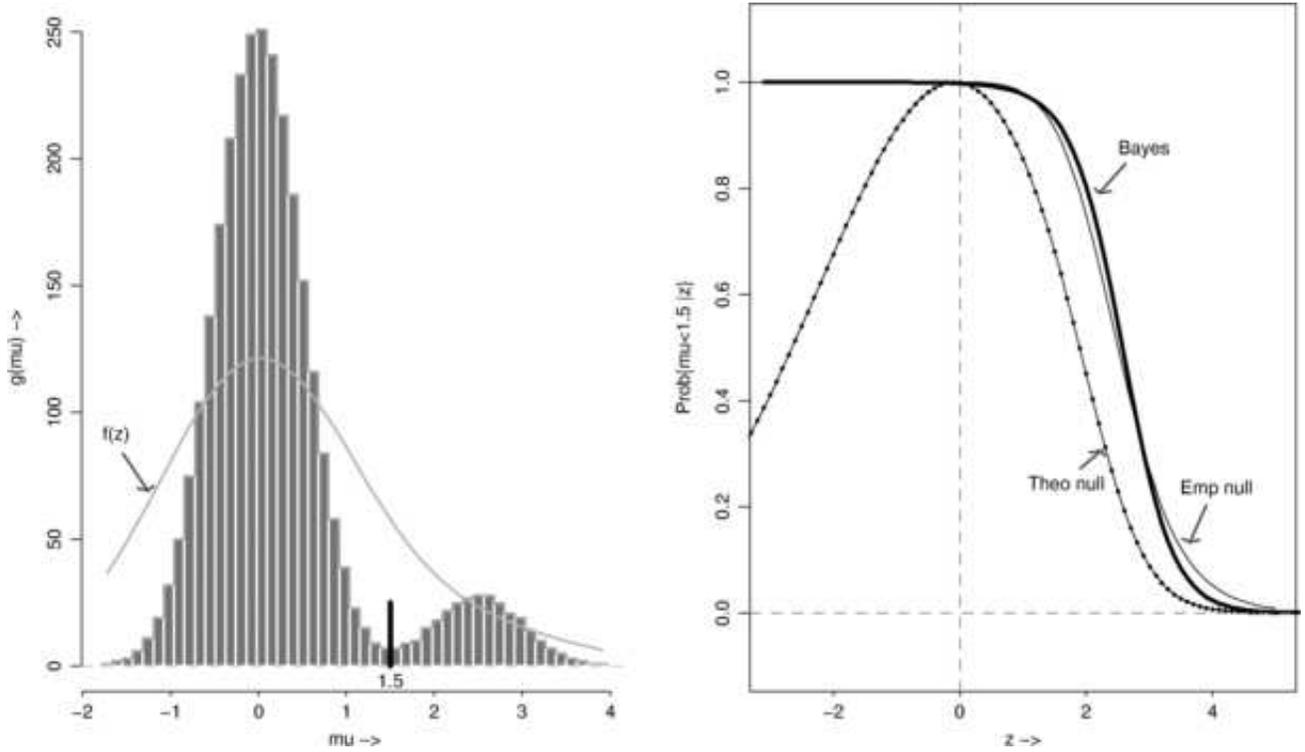

FIG. 7. *Left panel: Histogram shows $N = 3000$ draws of $\mu_i$ from model (6.2); smooth curve is corresponding density $f(z)$, (6.3). Right panel: "Emp null" is $\mathrm{fdr}(z)$ based on empirical null; it closely matches full Bayes posterior probability "Bayes" $= \Pr\{\mu_k < 1.5 | z\}$ from (6.1)–(6.2); "Theo null" is $\mathrm{fdr}(z)$ based on theoretical null, a poor match to Bayes.*

The density of $z$ in model (6.1) is

$$f(z) = \int_{-\infty}^{\infty} \varphi(\mu - z) g(\mu) \, d\mu,$$
(6.3)
$$[\varphi(x) = \exp(-x^2/2)/\sqrt{2\pi}],$$

shown as the smooth curve in the left-hand panel,

(6.4) $\quad f(z) = 0.9 \cdot \varphi_{0, 1.12}(z) + 0.1 \cdot \varphi_{2.5, 1.12}(z).$

The effect of noise in going from $\mu_i$ to $z_i \sim N(\mu_i, 1)$ has blurred the strongly bimodal $\mu$-histogram into a smoothly unimodal $f(z)$.

We can still employ the tactic of Figure 5, fitting a quadratic curve to $\log f(z)$ around $z = 0$ to estimate $p_0$ and the empirical null density $f_0(z)$. Using the formulas described later in this section gives

(6.5) $\quad p_0 = 0.93 \quad \text{and} \quad f_0(z) \sim N(.02, 1.14^2),$

and corresponding fdr curve $p_0 f_0(z)/f(z)$, labeled "Emp null" in the right-hand panel of Figure 7.

Looking at the histogram, it is reasonable to consider "interesting" those cases with $\mu_i \geq 1.5$, and "uninteresting" $\mu_i < 1.5$. The curve labeled "Bayes" in Figure 7 is the posterior probability $\Pr\{\text{uninteresting} | z\}$ based on full knowledge of (6.1), (6.2). The empirical null fdr curve provides an excellent estimate of the full Bayes result, without the prior knowledge. [An fdr based on the theoretical $N(0, 1)$ null is seen to be far off.]

Unobserved covariates, Reason 2 in Section 4, can easily produce blurry null hypotheses like that in (6.2). My point here is that the two-group model will handle blurry situations if the null hypothesis is empirically estimated. Or, to put things negatively, theoretical or permutation null methods are prone to error in such situations, no matter what kind of analysis technique is used.

Comparing (6.5) with (6.4) shows that $f_0(z)$ is just about right, but $p_0$ is substantially larger than the value 0.90 we might expect. The $\varphi_{2.5, .5}$ component of $g(\mu)$ puts some of its $z$-values near zero, weakening the zero assumption (4.1) and biasing $p_0$ upward. The same thing happened in Table 2 even though model (3.11) is "unblurred," $g(\mu)$ having a point mass at $\mu = 0$. Fortunately, $p_0$ is the least important part of the two-groups model for estimating $\mathrm{fdr}(z)$, under assumption (2.2). "Bias" can be a misleading term in model (6.1) since it presupposes that each $\mu_i$ is clearly defined as null or nonnull. This seems clear enough in (3.11). The null/nonnull distinction is less clear in (6.2), though it still makes sense to search for cases that have $\mu_i$ unusually far from 0.

The results in (6.5) come from a theoretical analysis of model (6.1). The idea in what follows is to generalize the construction in Figure 5 by approximating $\ell(z) = \log f(z)$ with Taylor series other than quadratic.

The $J$th Taylor approximation to $\ell(z)$ is

$$\ell_J(z) = \sum_{j=0}^{J} \ell^{(j)}(0) z^j / j!, \tag{6.6}$$

where $\ell^{(0)}(0) = \log f(0)$ and for $j \geq 1$

$$\ell^{(j)}(0) = \left. \frac{d^j \log f(z)}{dz^j} \right|_{z=0}. \tag{6.7}$$

Let $\widetilde{f}_0(z)$ indicate the subdensity $p_0 f_0(z)$, the numerator of fdr$(z)$ in (2.7). The choice

$$\widetilde{f}_0(z) = e^{\ell_J(z)} \tag{6.8}$$

matches $f(z)$ at $z=0$ (a convenient form of the zero assumption) and leads to an fdr expression

$$\mathrm{fdr}(z) = e^{\ell_J(z)} / f(z). \tag{6.9}$$

Larger choices of $J$ match $\widetilde{f}_0(z)$ more accurately to $f(z)$, increasing ratio (6.9); the interesting $z$-values, those with small fdr's, are pushed farther away from zero as we allow more of the data structure to be explained by the null density.

Bayesian model (6.1) provides a helpful interpretation of the derivatives $\ell^{(j)}(0)$:

LEMMA. *The derivative $\ell^{(j)}(0)$, (6.7), is the $j$th cumulant of the posterior distribution of $\mu$ given $z = 0$, except that $\ell^{(2)}(0)$ is the second cumulant minus 1. Thus*

$$\begin{aligned} \ell^{(1)}(0) &= E_0 \quad and \\ -\ell^{(2)}(0) &= 1 - V_0 \equiv \bar{V}_0, \end{aligned} \tag{6.10}$$

*where $E_0$ and $V_0$ are the posterior mean and variance of $\mu$ given $z=0$.*

Proof of the lemma appears in Section 7 of Efron (2005).

For $J = 0, 1, 2$, formulas (6.8), (6.9) yield simple expressions for $p_0$ and $f_0(z)$ in terms of $f(0)$, $E_0$ and $\bar{V}_0$. These are summarized in Table 4, with $p_0$ obtained from

$$p_0 = \int_{-\infty}^{\infty} \widetilde{f}_0(z) \, dz. \tag{6.11}$$

TABLE 4
*Expressions for $p_0$, $f_0$ and fdr, first three choices of $J$ in (6.8), (6.9); $\bar{V}_0 = 1 - V_0$; $J=0$ gives theoretical null, $J=2$ empirical null; $f(z)$ from (6.3)*

| J | 0 | 1 | 2 |
|---|---|---|---|
| $p_0$ | $f(0)\sqrt{2\pi}$ | $f(0)\sqrt{2\pi} e^{E_0^2/2}$ | $f(0)\sqrt{\frac{2\pi}{\bar{V}_0}} e^{E_0^2/2\bar{V}_0}$ |
| $f_0(z)$ | $N(0,1)$ | $N(E_0, 1)$ | $N(E_0/\bar{V}_0, 1/\bar{V}_0)$ |
| fdr$(z)$ | $\frac{f(0) e^{-z^2/2}}{f(z)}$ | $\frac{f(0) e^{E_0 z - z^2/2}}{f(z)}$ | $\frac{f(0) e^{E_0 z - \bar{V}_0 z^2/2}}{f(z)}$ |

Formulas are also available for Fdr$(z)$, (2.8).

The choices $J=0,1,2$ in Table 4 result in a normal null density $f_0(z)$, the only difference being the means and variances. Going to $J=3$ allows for an asymmetric choice of $f_0(z)$,

$$\mathrm{fdr}(z) = \frac{f(0)}{f(z)} e^{E_0 z - \bar{V}_0 z^2/2 + S_0 z^3/6}, \tag{6.12}$$

where $S_0$ is the posterior third central moment of $\mu$ given $z=0$ in model (6.1). The program *locfdr* uses a variant, the "split normal," to model asymmetric null densities, with the exponent of (6.12) replaced by a quadratic spline in $z$.

The lemma bears on the difference between empirical and theoretical nulls. Suppose that the probability mass of $g(\mu)$ occurring within a few units of the origin is concentrated in an atom at $\mu = 0$. Then the posterior mean and variance $(E_0, V_0)$ of $\mu$ given $z = 0$ will be near 0, making $(E_0, \bar{V}_0) \doteq (0,1)$. In this case the empirical null $(J=2)$ will approximate the theoretical null $(J=0)$. Otherwise the two nulls differ; in particular, any mass of $g(\mu)$ near zero increases $V_0$, swelling the standard deviation $(1-V_0)^{-1/2}$ of the empirical null.

The two-groups model (2.1), (2.2) puts one in a hypothesis-testing frame of mind: a large group of uninteresting cases is to be statistically separated from a small interesting group. Even blurry situations like (6.2) exhibit a clear grouping, as in Figure 7. None of this is necessary for the one-group model (6.1). We might, for example, suppose that $g(\mu)$ is normal,

$$\mu \sim N(A, B^2), \tag{6.13}$$

and proceed in an empirical Bayes way to estimate $A$ and $B$ and then apply Bayes estimation to the individual cases.

This line of thought leads directly to James–Stein estimation (Efron and Morris, 1975). Estimation, as opposed to testing, is the key word here—with possible effect sizes $\mu_i$ varying continuously rather than having a large clump of values near zero. The Education data of panel B, Figure 1, could reasonably be analyzed this way, instead of through simultaneous testing. Scientific context, which says that there is likely to be a large group of (nearly) unaffected genes, as in (2.2), is what makes the two-groups model a reasonable Bayes prior for microarray studies.

## 7. BAYESIAN AND FREQUENTIST CONFIDENCE STATEMENTS

False discovery rate methods provide a happy marriage between Bayesian and frequentist approaches to multiple testing, as shown in Section 2. Empirical Bayes techniques based on the two-groups model seem to give us the best of both statistical philosophies. Things do not always work out so peaceably; in these next two sections I want to discuss contentious situations where the divorce court looms as a possibility.

An insightful and ingenious paper by Benjamini and Yekutieli (2005) discusses the following problem in simultaneous significance testing: having applied false discovery rate methods to select a set of nonnull cases, how can confidence intervals be assigned to the true effect size for each selected case? (The paper and the ensuing discussion are much more general, but this is all I need for the illustration here.)

Figure 8 concerns Benjamini and Yekutieli's solution applied to the following simulated data set: $N = 10{,}000$ $(\mu_i, z_i)$ pairs were generated as in (6.1), with 90% of the $\mu_i$ zero, the null cases, and 10% distributed $N(-3, 1)$,

$$(7.1) \quad g(\mu) = 0.90 \cdot \delta_0(\mu) + 0.10 \cdot \varphi_{-3,1}(\mu),$$

$\delta_0(\mu)$ a delta function at $\mu = 0$. The Fdr procedure (2.5) was applied with $q_0 = 0.05$, yielding 566 nonnull "discoveries," those having $z_i \leq -2.77$.

The Benjamini–Yekutieli "false coverage rate" (FCR) control procedure provides upper and lower bounds for the true effect size $\mu_i$ corresponding to each $z_i$ less than $-2.77$; these are indicated by heavy diagonal lines in Figure 8, constructed as described in BY's Definition 1. This construction guarantees that the expected proportion of the 566 intervals *not* containing the true $\mu_i$, the false coverage rate, is bounded by $q = 0.05$.

In a real application only the $z_i$'s and their BY confidence intervals could be seen, but in a simulation we can plot the actual $(z_i, \mu_i)$ pairs, and compare them to the intervals. Figure 8 plots $(z_i, \mu_i)$ for the 1000 nonnull cases, those from $\mu_i \sim N(-3, 1)$ in (7.1). Of these, 55,2 plotted as heavy points, lie to the left of $z_0 = -2.77$, the Fdr threshold, with the other 448 plotted as light points; 14 null cases, $\mu_i = 0$, plotted as "+," also had $z_i < z_0$.

The first thing to notice is that the FCR property is satisfied: only 17 of the 566 intervals have failed to contain $\mu_i$ (14 of these the +'s), giving 3% noncoverage. The second thing, though, is that the intervals are frighteningly wide—$z_i \pm 2.77$, about $\sqrt{2}$ longer than the usual individual 95% intervals $z_i \pm 1.96$—and poorly centered, particularly at the left where all the $\mu_i$'s fall in their intervals' upper halves.

An interesting comparison is with Bayes' rule applied to (6.1), (7.1), which yields

$$(7.2) \qquad \Pr\{\mu = 0 | z_i\} = \mathrm{fdr}(z_i),$$

where

$$(7.3) \quad \begin{aligned} \mathrm{fdr}(z) &= 0.9 \cdot \varphi_{0,1}(z) \\ &\quad \cdot [0.9 \cdot \varphi_{0,1}(z) + 0.1 \cdot \varphi_{-3,\sqrt{2}}(z)]^{-1} \end{aligned}$$

as in (2.7), and

$$(7.4) \qquad g(\mu_i | \mu_i \neq 0, z_i) \sim N\left(\frac{z_i - 3}{2}, \frac{1}{2}\right).$$

That is, $\mu_i$ is null with probability $\mathrm{fdr}(z_i)$, and $N((z_i - 3)/2, 1/2)$ with probability $1 - \mathrm{fdr}(z_i)$. The dashed lines indicate the posterior 95% intervals given that $\mu_i$ is nonnull, $(z_i - 3)/2 \pm 1.96/\sqrt{2}$, now $\sqrt{2}$ *shorter* than the usual individual intervals; at the top of Figure 9 the beaded curve shows $\mathrm{fdr}(z_i)$.

The frequentist FCR intervals and the Bayes intervals are pursuing the same goal, to include the nonnull scores $\mu_i$ with 95% probability. At $z_i = -2.77$ the FCR assessment is $\Pr\{\mu \in [-5.54, 0]\} = 0.95$; Bayes' rule states that $\mu_i = 0$ with probability $\mathrm{fdr}(-2.77) = 0.25$, and if $\mu_i \neq 0$, then $\mu_i \in [-4.27, -1.49]$ with probability 0.95. This kind of disconnected description is natural to the two-groups model. A principal cause of FCR's oversized intervals (the paper shows that no FCR-controlling intervals can be much narrower) comes from using a single connected set to describe a disconnected situation.

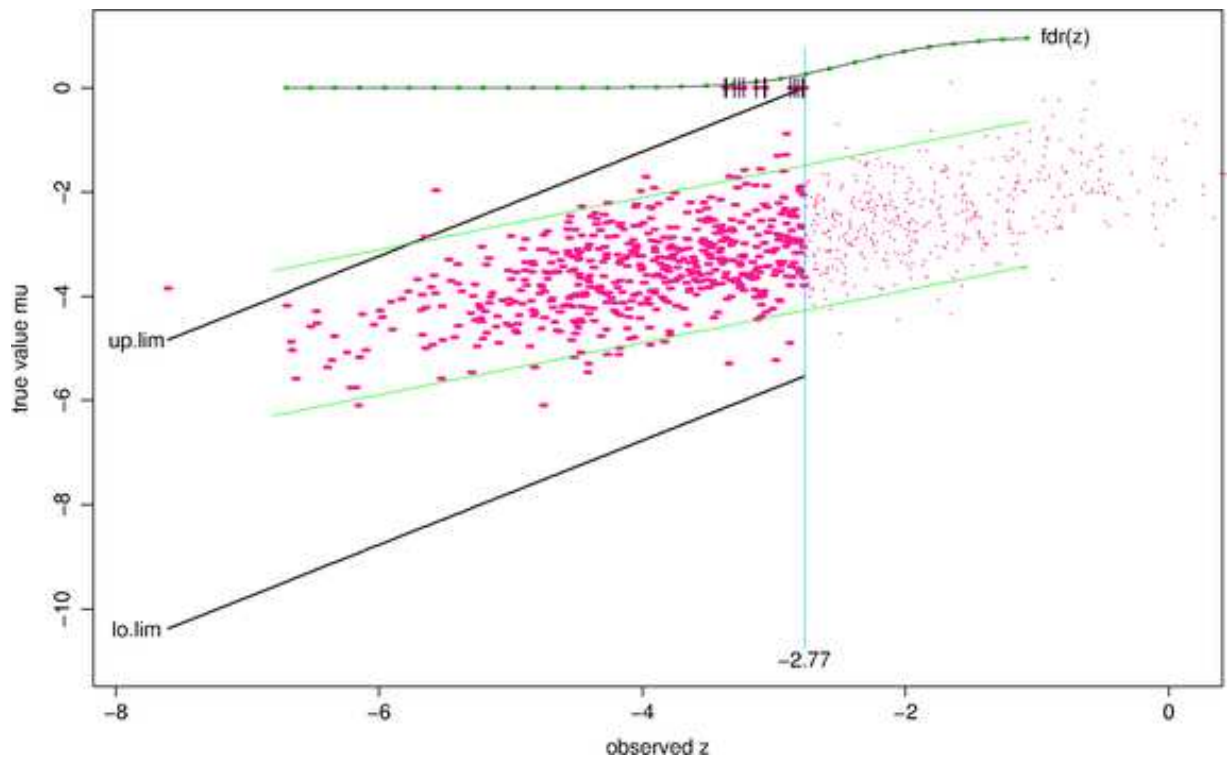

FIG. 8. *Benjamini–Yekutieli FCR controlling intervals applied to simulated sample of 10,000 cases from (6.1), (7.1). 566 cases have $z_i \leq z_0 = -2.77$, the Fdr (0.05) threshold. Plotted points are $(z_i, \mu_i)$ for the 1000 nonnull cases; 14 null cases with $z_i \leq z_0$ indicated by "+." Heavy diagonal lines indicate FCR 95% interval limits; light lines are Bayes 95% posterior intervals given $\mu_i \neq 0$. Beaded curve at top is $fdr(z_i)$, posterior probability $\mu_i = 0$.*

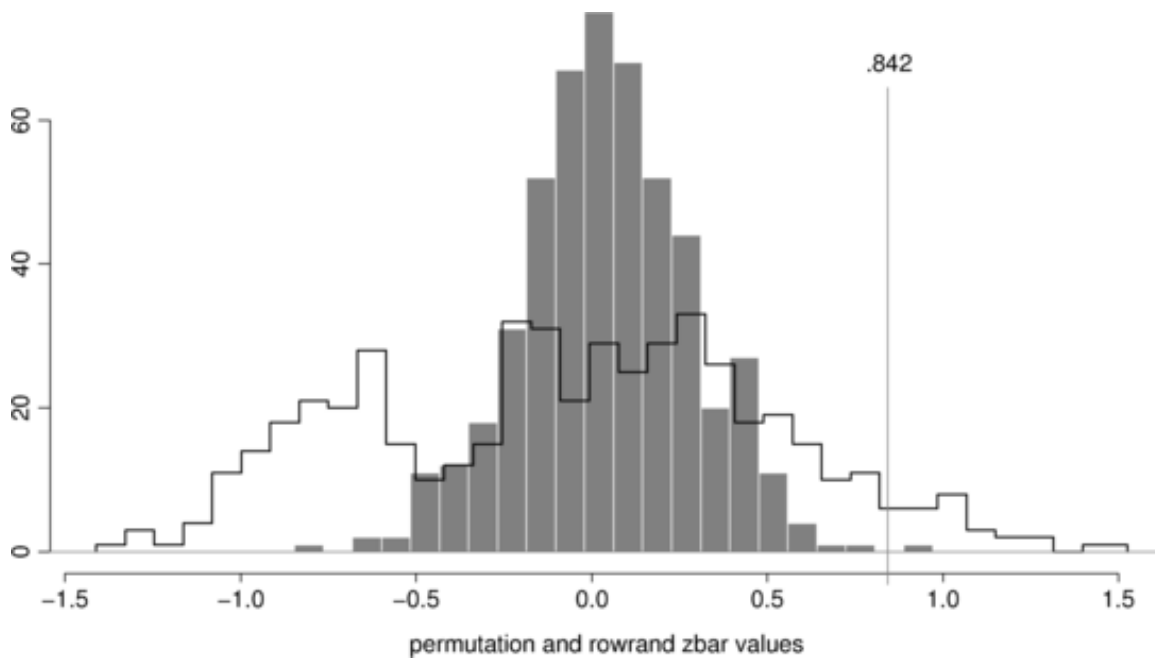

FIG. 9. *Computing a p-value for $\bar{z}_S = 0.842$, average of 15 z-values in CTL pathway, p53 data Solid histogram 500 row randomizations give p-value 0.002. Line histogram 500 column permutations give p-value 0.048.*

Of course Bayes' rule will not be easily available to us in most practical problems. Is there an empirical Bayes solution? Part of the solution certainly is there: estimating fdr($z$) as in Section 3. Estimating $g(\mu_i|\mu_i \neq 0, z_i)$, (7.4), is more challenging. A straightforward approach uses the nonnull counts (3.8) to estimate the nonnull density $f_1(z)$ in (2.1), deconvolutes $\widehat{f}_1(z)$ to estimate the nonnull component "$g_1(\mu)$" in (7.1), and applies Bayes' rule directly to $\widehat{g}_1$. This works reasonably well in Figure 8's example, but deconvolution calculations are notoriously tricky and I have not been able to produce a stable general algorithm.

Good frequentist methods like the FCR procedure enjoy the considerable charm of an exact error bound, without requiring a priori specifications, and of course there is no law that they have to agree with any particular Bayesian analysis. In large-scale situations, however, empirical Bayes information can overwhelm both frequentist and Bayesian predilections, hopefully leading to a more satisfactory compromise between the two sets of intervals appearing in Figure 8.

## 8. IS A SET OF GENES ENRICHED?

Microarray experiments, through a combination of insufficient data per gene and massively multiple simultaneous inference, often yield disappointing results. In search of greater detection power, *enrichment analysis* considers the combined outcomes of biologically defined sets of genes, such as pathways. As a hypothetical example, if the 20 $z$-values in a certain pathway all were positive, we might infer significance to the pathway's effect, whether or not any of the individual $z_i$'s were deemed nonnull.

Our example here will involve the *p53 data*, from Subramanian et al. (2005), $N = 10{,}100$ genes on $n = 50$ microarrays, $z_i$'s as in (3.2), whose $z$-value histogram looks like a slightly short-tailed normal distribution having mean 0.04 and standard deviation 1.06. Fdr analysis (2.5), $q = 0.1$, yielded just one nonnull gene, while enrichment analysis indicated seven or eight significant gene sets, as discussed at length in Efron and Tibshirani (2006).

Figure 9 concerns the *CTL pathway*, a set of 15 genes relating to the development of so-called killer T cells, #95 in a catalogue of 522 gene-sets provided by Subramanian et al. (2005). For a given gene-set "$\mathcal{S}$" with $m$ members, let $\bar{z}_{\mathcal{S}}$ denote the mean of the $m$ $z$-values within $\mathcal{S}$; $\bar{z}_{\mathcal{S}}$ is the enrichment statistic suggested in the Bioconductor $R$ package *limma* (Smyth, 2004),

$$(8.1) \qquad \bar{z}_{\mathcal{S}} = 0.842$$

for the CTL pathway. How significant is this result? I will consider assigning an individual $p$-value to (8.1), not taking into account multiple inference for a catalogue of possible gene-sets (which we could correct for later using Fdr methods, for instance, to combine the individual $p$-values).

Limma computes $p$-values by "row randomization," that is, by randomizing the order of rows of the $N \times n$ expression matrix $X$, and recomputing the statistic of interest. For a simple average like (8.1) this amounts to choosing random subsets of size $m = 15$ from the $N = 10{,}100$ $z_i$'s and comparing $\bar{z}_{\mathcal{S}}$ to the distribution of the randomized values $\bar{z}_{\mathcal{S}}^*$. Five hundred rowrands produced only one $\bar{z}_{\mathcal{S}}^* > \bar{z}_{\mathcal{S}}$, giving $p$-value $1/500 = 0.002$.

Subramanian et al. calculate $p$-values by permuting the *columns* of $X$ rather than the rows. The permutations yield a much wider distribution than the row randomizations in Figure 9, with corresponding $p$-value 0.048. The reason is simple: the genes in the CTL pathway have highly correlated expression levels that increase the variance of $\bar{z}_{\mathcal{S}}^*$; column-wise permutations of $X$ preserve the correlations across genes, while row randomizations destroy them.

At this point it looks like column permutations should always give the right answer. Wrong! For the BRCA data in Figure 4, the ensemble of $z$-values has (mean, standard deviation) about $(0, 1.50)$, compared to $(0, 1)$ for $z_i^*$'s from column permutations. This shrinks the permutation variability of $\bar{z}_{\mathcal{S}}^*$, compared to what one would get from a random selection of genes for $\mathcal{S}$, and can easily reverse the relationship in Figure 9.

The trouble here is that there are two obvious, but different, null hypotheses for testing enrichment:

*Randomization null hypothesis* $\mathcal{S}$ has been chosen by random selection of $m$ genes from the full set of $N$ genes.

*Permutation null hypothesis* The order of the $n$ microarrays has been chosen at random with respect to the patient characteristics (e.g., with the patient being in the normal or cancer category in Example A of the Introduction).

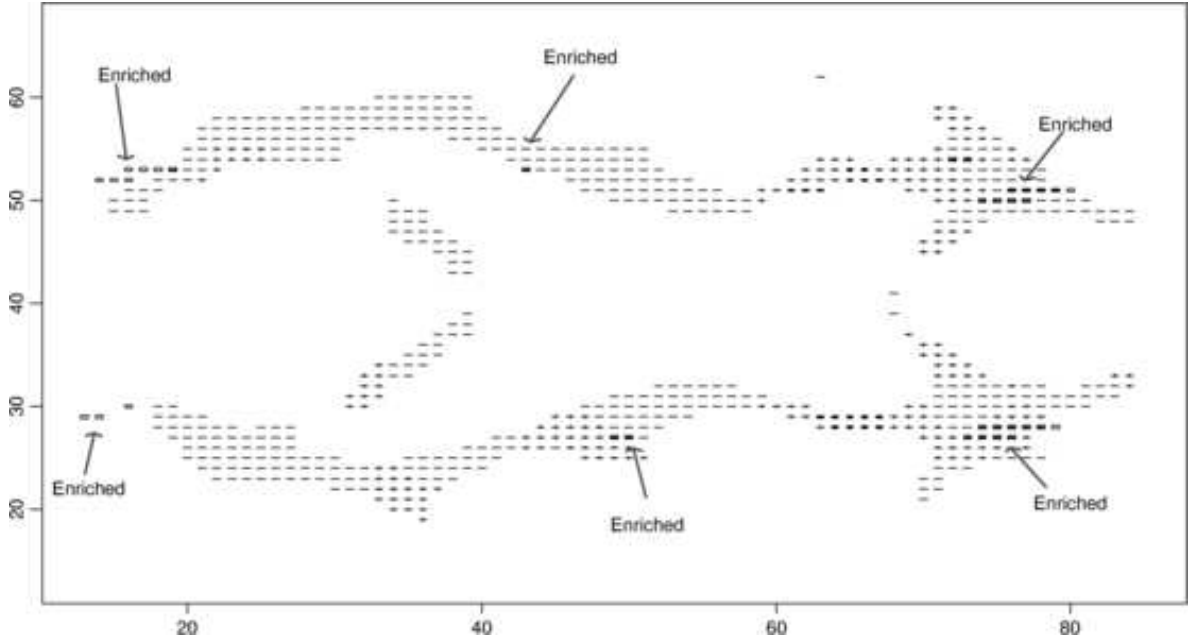

Fig. 10. *Enrichment analysis of Imaging data, panel* D *of Figure 1; z-value for original 15,445 voxels have been averaged over "gene-sets" of neighboring voxels with city-block distance $\leq 2$. Coded as "$-$" for $\bar{z}_i < 0$, "$+$" for $\bar{z}_i \geq 0$; solid rectangles, labeled as "Enriched," show voxels with $\widehat{\mathrm{fdr}}(\bar{z}_i) \leq 0.2$, using empirical null.*

Efron and Tibshirani (2006) suggest a compromise method, *restandardization*, that to some degree accommodates both null hypotheses. Instead of permuting $\bar{z}_{\mathcal{S}}$ in (8.1), restandardization permutes $(\bar{z}_{\mathcal{S}} - \mu_z)/\sigma_z$, where $(\mu_z, \sigma_z)$ are the mean and standard deviation of all $N$ $z_i$'s. Subramanian et al. do something similar using a Kolmogorov–Smirnov enrichment statistic.

All of these methods are purely frequentistic. Theoretically we might consider applying the two-groups/empirical Bayes approach to sets of $z$-values "$\mathbf{z}_{\mathcal{S}}$," just as we did for individual $z_i$'s in Sections 2 and 3. For at least three reasons that turns out to be extremely difficult:

- My technique for estimating the mixture density $f$, as in (3.6), becomes exponentially more difficult in higher dimensions.
- There is not likely to be satisfactory theoretical null $f_0$ for the correlated components of $\bar{z}_{\mathcal{S}}$, while estimating an empirical null faces the same "curse of dimensionality" as for $f$.
- As discussed following (3.10), false discovery rate interpretation depends on exchangeability, essentially an equal a priori interest in all $N$ genes. There may be just one gene-set $\mathcal{S}$ of interest to an investigator, or a catalogue of several hundred $\mathcal{S}$'s as in Subramanian et al., but we certainly are not interested in all possible gene-sets. It would be a daunting exercise in subjective, as opposed to empirical, Bayesianism to assign prior probabilities to any particular gene-set $\mathcal{S}$.

Having said this, it turns out there is one "gene-set" situation where the two-groups/empirical Bayes approach is practical (though it does not involve genes). Looking at panel D of Figure 1, the Imaging data, the obvious spatial correlation among $z$-values suggests local averaging to reduce the effects of noise.

This has been carried out in Figure 10: at voxel $i$ of the $N = 15,445$ voxels, the average of $z$-values for those voxels within city-block distance 2 has been computed, say "$\bar{z}_i$." The results for the same horizontal slice as in panel D are shown using a similar symbol code. Now that we have a single number $z_i$ for each voxel, we can compute the empirical null $\widehat{\mathrm{fdr}}$ estimates as in Section 4. The voxels labeled "enriched" in Figure 10 are those having $\widehat{\mathrm{fdr}}(\bar{z}_i) \leq 0.2$.

Enrichment analysis looks much more familiar in this example, being no more than local spatial smoothing. The convenient geometry of three-dimensional space has come to our rescue, which it emphatically fails to do in the microarray context.

## 9. CONCLUSION

Three forces influence the state of statistical science at any one time: mathematics, computation and applications, by which I mean the type of problems subject-area scientists bring to us for solution. The Fisher–Neyman–Pearson theory of hypothesis testing was fashioned for a scientific world where experimentation was slow and difficult, producing small data sets focused on answering single questions. It was wonderfully successful within this milieu, combining elegant mathematics and limited computational equipment to produce dependable answers in a wide variety of application areas.

The three forces have changed relative intensities recently. Computation has become literally millions of times faster and more powerful, while scientific applications now spout data in fire-hose quantities. (Mathematics, of course, is still mathematics.) Statistics is changing in response, as it moves to accommodate massive data sets that aim to answer thousands of questions simultaneously. Hypothesis testing is just one part of the story, but statistical history suggests that it could play a central role: its development in the first third of the twentieth century led directly to confidence intervals, decision theory and the flowering of mathematical statistics.

I believe, or maybe just hope, that our new scientific environment will also inspire a new look at old philosophical questions. Neither Bayesians nor frequentists are immune to the pressures of scientific necessity. Lurking behind the specific methodology of this paper is the broader, still mainly unanswered, question of how one should combine evidence from thousands of parallel but not identical hypothesis testing situations. What I called "empirical Bayes information" accumulates in a way that is not well understood yet, but still has to be acknowledged: in the situations of Figure 4, the frequentist is not free to stick with classical null hypotheses, while the Bayesian cannot use prior (6.13), at least not without the risk of substantial inferential confusion.

Classical statistics developed in a data-poor environment, as Fisher's favorite description, "small-sample theory," suggests. By contrast, modern-day disciplines such as machine learning seem to struggle with the difficulties of too much data. Both problems, too little and too much data, can afflict microarray studies. Massive data sets like those in Figure 1 are misleadingly comforting in their suggestion of great statistical accuracy. As I have tried to show here, the power to detect interesting specific cases, genes, may still be quite low. New methods are needed, perhaps along the lines of "enrichment," as well as a theory of experimental design explicitly fashioned for large-scale testing situations.

One floor up from the philosophical basement lives the untidy family of statistical models. In this paper I have tried to minimize modeling decisions by working directly with $z$-values. The combination of the two-groups model and false discovery rates applied to the $z$-value histogram is notably light on assumptions, more so when using an empirical null, which does not even require independence across the columns of $X$ (i.e., across microarrays, a dangerous assumption as shown in Section 6 of Efron, 2004). There will certainly be situations when modeling inside the $X$ matrix, as in Newton et al. (2004) or Kerr, Martin and Churchill (2000), yields more information than $z$-value procedures, but I will leave that for others to discuss.

## ACKNOWLEDGMENTS

This research was supported in part by the National Science Foundation Grant DMS-00-72360 and by National Institute of Health Grant 8R01 EB002784.